\RequirePackage{fix-cm}
\documentclass[twocolumn]{svjour3} 

\smartqed 
\usepackage{graphics}
\usepackage[lined, linesnumbered, ruled]{algorithm2e}
\usepackage[cmex10]{amsmath}
\usepackage{amssymb}
\usepackage{subfigure}
\usepackage{multirow}
\usepackage{epstopdf}
\usepackage{wrapfig}
\usepackage{rotating}
\usepackage{morefloats}
\usepackage[acronym,nonumberlist]{glossaries}

\usepackage[numbers,sort]{natbib}
\begin{document}
\title{Adaptive Multi-polling Scheduler for QoS Support of Video Transmission in IEEE 802.11e WLANs
}


\author{Mohammed~A.~Al-Maqri \and
Mohamed Othman \and
Borhanuddin Mohd Ali \and
Zurina Mohd Hanapi
}


\institute{Mohammed A. Al-Maqri and Mohamed Othman\at
Department of Communication Technology and Network, Universiti Putra Malaysia, Malaysia \\
mohdalmoqry@gmail.com (M. A. Al-Maqri), \\ mothman@upm.edu.my (M. Othman) 
}

\date{Received: date / Accepted: date}

\maketitle
\sloppy
\begin{abstract}
The 802.11E Task Group has been established to enhance Quality of Service (QoS) provision for time-bounded services in the current IEEE 802.11 Medium Access Control (MAC) protocol. The QoS is introduced throughout Hybrid Coordination Function Controlled Channel Access (HCCA) for the rigorous QoS provision. In HCCA, the station is allocated a fixed Transmission Opportunity (TXOP) based on its TSPEC parameters so that it is efficient for Constant Bit Rate streams. However, as the profile of Variable Bit Rate (VBR) traffics is inconstant, they are liable to experience a higher delay especially in bursty traffic case. In this paper, we present a dynamic TXOP assignment algorithm called Adaptive Multi-polling TXOP scheduling algorithm (AMTXOP) for supporting the video traffics transmission over IEEE 802.11e wireless networks. This scheme invests a piggybacked information about the size of the subsequent video frames of the uplink streams to assist the Hybrid Coordinator accurately assign the TXOP according to actual change in the traffic profile. The proposed scheduler is powered by integrating multi-polling scheme to further reduce the delay and polling overhead. Extensive simulation experiments have been carried out to show the efficiency of the AMTXOP over the existing schemes in terms of the packet delay and the channel utilization.
\keywords{Quality of Service; WLAN; IEEE 802.11e; HCCA; TXOP; Multi-polling}
\end{abstract}

\section{Introduction}
\label{sec:intro}
Recent years have witnessed a rapid growth of ubiquitous applications in the Internet with a vast spread of multimedia streams. This makes providing differentiated Quality of Service (QoS) for such applications in Wireless Local Area Networks (WLANs) very challenging task. Besides, several wireless technologies have been risen from amongst them IEEE 802.11 \cite{IEEEStand1999} has assumed as a de facto standard in WLANs due to some of its key features like deployment flexibility, infrastructure simplicity and cost effectiveness \cite{Ammar2011}. In IEE 802.11 WLANs, the QoS of multimedia communications cannot be efficiently achieved due to frequent collisions and retransmission \cite{Abouaissa2013}. IEEE 802.11 introduces two channel access modes, namely Distributed Coordination Function (DCF) and Point Coordination Function (PCF). The former is the mandatory medium access method which is appropriate to serve best effort applications such as HTTP, FTP and SMTP. Multimedia streams that require a certain QoS level are served during the controlled mode (i.e. PCF) since it provides a contention-free polling-based access to the channel to provide the demanded QoS. However, due to the fact that PCF only operates on the Free-Contention period, which may considerably cause an increase in the transmission delay, especially with high bursty traffics it considered not efficient for serving the applications that required high QoS constraints. Therefore, IEEE 802.11 Task Group e (TGe) has established IEEE 802.11e protocol \cite{IEEEStandard2007} which then introduced a revised version \cite{IEEEStandard2012} with new technical enhancements on MAC and Physical layer.

The QoS support of IEEE 802.11 standard has been extended in IEEE 802.11e by means of Hybrid Coordination Function (HCF). Enhanced Distributed Channel Access function (EDCF) which extends DCF, provides a prioritized QoS throughout its distributed access manner to the wireless medium. HCF Controlled Channel Access (HCCA) which extends PCF that works based on a centralized polling mechanism to provide differentiated service, according to rigid QoS parameters negotiated with the centralized coordination (HC). EDCA introduces a random access to the wireless medium by means of access categories (ACs). The traffics are mapped to ACs according to their priority. Every AC will be associated with a backoff timer so that the highest priority ACs will go through a shorter backoff process. Despite EDCA provides QoS support, it is still not efficient for application with rigid QoS requirements. 

The QoS level required by the delay-sensitive application can be better met by HCCA since it is designed to minimize the control messaging overhead imposed by the distributed approach of EDCF. Besides, strict QoS support of HCCA can be exploited to transmit some TCP-based services such as web-enabled YouTube progressive video download \cite{navarro2013}. Indeed, provisioning QoS of CBR applications such as G.711 \cite{G7111988} audio streams and (H.261/MPEG-1) video \cite{MPEG11997} is suitable. Whereas, HCCA fails to support VBR traffics based on their declared TSPEC parameters as they show irregularity in their profile during the time.
Two-third of all traffics in the networks will be video by 2017 according to Cisco Visual Networking Index \cite{indexglobal}. This motivates many researches to be carried out to improve the QoS provisioning in WLANs in terms of supporting QoS for such streams. Recently, Motion Picture Experts Group type 4 (MPEG--4/H.264) has become a prominent video the internet due to its scalability, error robustness and network-friendly features \cite{Mao2007}. In HCCA, scheduling VBR streams with fixed TXOP throughout their lifetime can provide sufficient yet not favorable service quality as the data rate of such streams are likely varied from its mean characteristics during the time. This can lead to a noticeable increase in the packet delay and deterioration in channel utilization.

More recently, a large deal of video consumed on mobile devices has been prerecorded video which is considered non-realtime video \cite{Gautam2013}. The pre-recorded videos can be utilized in several applications and forms such as streaming of video surveillance, providing a large amount of access to knowledge in educational systems \cite{chaudhuri2013} and in peer-to-peer (P2P) streaming where users can share stored their video content \cite{birkos2013}. Besides, the spreading of User-Generated Content (UGC) for instance, pre-recorded/stored videos such as YouTube, Dailymotion, Metacafe have become noticeable in Internet. In the stored video, the entire traffic can be precisely identified to allocate an appropriate amount of bandwidth so as to keep the cost functions (such as packet delay) minimized. In wireless infrastructure environments the scheduling of the uplink stream considered more complicated than downlink streams as the coordinator does not aware about the queue status of the terminals \cite{Teixeira2013}.
Motivated by the widespread of pre-recorded videos in nowadays wireless networks and the availability of the entire video traffic behavior prior to the traffic setup, we introduce a new adaptive scheduler. This scheduler exploits the availability of pre-recorded video profile in order to leverage the scheduling process.
To the best of our knowledge, the scheduling of uplink stored continuous media in HCCA has not been addressed efficiently in WLANs. A novel scheduler is presented in this paper aiming at adapt to the fast changes of VBR video traffic profile. Some researches have highlighted this fact, such as that in \cite{Panos2002,Fang2006raey,Haddad2012}. The proposed scheduler utilizes the availability of the entire characteristics of pre-recorded streams for scheduling purposes. It computes the TXOP for a traffic stream based on knowledge about the actual frame size instead of assigning TXOP according to mean characteristics of the traffic which is unable to reflect the actual traffic. This scheduler exploits the Queue Size field (QS) of IEEE 802.11e MAC header to transfer this information to the HC. Furthermore, since the poll message can be overheard by all QSTAs within the BSS, broadcasting one multiple poll message to all QSTAs in the SI rather than send one poll for each will minimize the poll overhead and the packet delay with improving the bandwidth utilization. Hence, to leverage the benefit of the proposed algorithm a multi-polling feature has been integrated. We exploit similar multi-polling frame to that reported in \cite{ByungSeoKim2005} and \cite{Chou2011} which is illustrated in details in sub-section~\ref{sec:MPFS}.

The rest of the paper is organized as follows. Section ~\ref{sec:relatedWorks} illustrates the reference HCCA mechanism and its deficiency in supporting VBR Video streams and demonstrates some of the HCCA related works. Section ~\ref{sec:ATAV} illustrates the proposed dynamic algorithm. The performance evaluation including simulation and analytical results is discussed in Section ~\ref{sec:evaluation}. Section ~\ref{sec:conclusion} concludes the work presented in this paper.

\section{Background and Related Works}
\label{sec:relatedWorks}
This section presents a description of IEEE 802.11e HCCA scheduler and some MPEG--4 video characteristics. The deficiency of HCCA in supporting VBR is illustrated and some related works in enhancing its performance are also discussed.

\subsection{IEEE 802.11e HCCA Mechanism}
\begin{figure*}
\centering
\includegraphics[width=\linewidth]{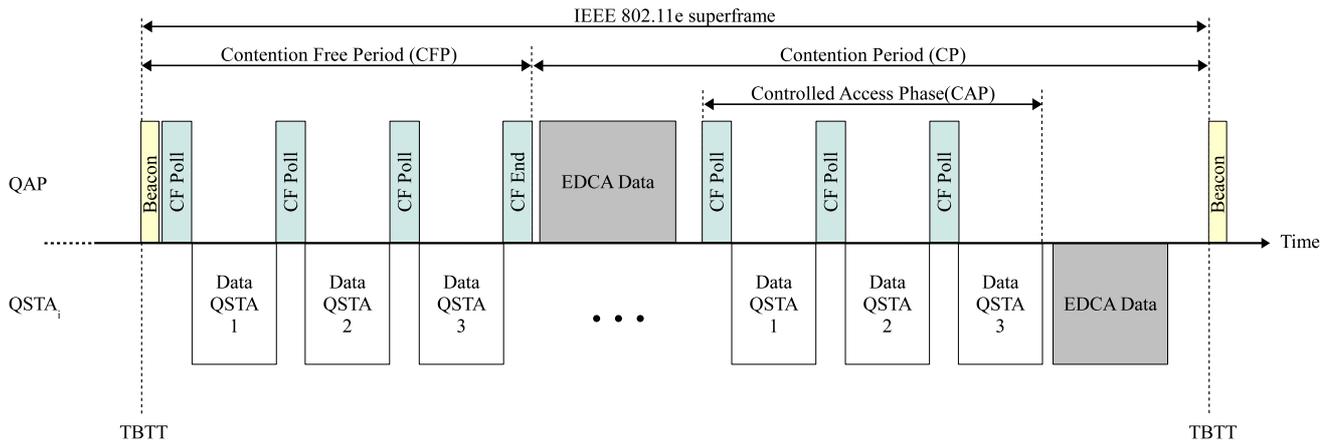}
\caption{Controlled channel access mechanism in IEEE 802.11e HCCA.}
\label{fig001}
\end{figure*}

In IEEE 802.11e, the channel divided into occurrences of subsequent superframe which includes Free Period (CFP) and Contention Period (CP). CFP is reserved for centralized transmission, which controlled by HCCA function. 
However, Controlled Access Phase (CAP) can be initiated during CFP and CP as well when the medium remains idle for at lease a PIFS time. This property is one of the key merits of HCCA that make HCCA overcome the legacy PCF function in that it shorten the time the packet waits in its transmission queue. 
The scheduling process of QSTAs is discussed in details in the next section. Figure~\ref{fig001} demonstrates an example of HCCA transmission during CFP and CP periods.

\subsection{IEEE 802.11e HCCA scheduler}
\label{sec:HCCA}
In order to initiate an uplink traffic, the QSTA issues a QoS reservation through transmitting an ADDTS-Request frame. This frame carries information about the TSPEC which is required by HC for scheduling purpose. The list of mandatory fields of the TSPEC is described in \cite{Grilo2003}.
The HC which usually resides in the QoS-enabled Access Point (QAP) which maintains the TSPECs of all \textit{TSs} in the polling list. The HC computes the TXOP for each QSTA. 

Upon receiving an ADDTS-Request from a QSTA, the HCCA scheduler of the HC goes through the following steps to schedule the uplink traffics:
\begin{enumerate}
\item \textit{SI Assignment}

\label{eq:SIassign}
The scheduler calculates the Service Interval ($SI$) as the minimum of all Maximum Service Intervals ($MSI$) of all admitted traffic streams which is a submultiple of the beacon interval, the time between two subsequent TBTTs. The minimum $MSI$ for each QSTA is obtained from Equation~\eqref{eq:minSI}.
\begin{equation}
\label{eq:minSI}
MSI_{min} = min(MSI_{i}),\quad i =1, 2, 3, ..., n
\end{equation}
where $n$ is the number of admitted \textit{TSs} and $MSI_{i}$ is the maximum $SI$ of the $i^{th}$ stream.
The $SI$ is computed so that it satisfies the condition in Equation~\eqref{eq:si}.
\begin{equation}
\label{eq:si}
SI = \frac{BeaconInterval}{x} \leq MSI_{min},
\end{equation}
where the denominator, $x$, is an integer number that divides the beacon interval into the largest number that is equal or less than the $MSI_{min}$.\\
\item \textit{TXOP Allocation}

HC allocates a TXOP to each admitted QSTA so as to enable it to transmit its data with regards to the negotiated QoS parameters of the TSPEC.

For the $i^{th}$ QSTA, the scheduler computes the number of MSDUs arrived at the mean data rate $\rho_{i}$ as in Equation~\eqref{eq:n}.
\begin{equation}
N_{i} = \left \lceil \frac{SI \times \rho_{i}}{L_{i}} \right \rceil,
\label{eq:n}
\end{equation}
where $L_{i}$ is the nominal MSDU length for the $i^{th}$ QSTA.
Then the TXOP duration of the particular station, $TXOP_{i}$, is calculated as the maximum of the time required to transmit $N_{i}$ MSDU or the time to transmit one maximum MSDU at the minimum physical rate $R_{i}$, as stated in Equation~\eqref{eq:txop}.
\begin{equation}
TXOP_{i} = max\left (\frac{N_{i} \times L_{i}}{R_{i}}, \frac{M{i}}{R_{i}}\right) + O
\label{eq:txop}
\end{equation}
where $M$ is the maximum MSDU and $O$ denotes the overhead, including MAC and PHY headers, inter-frame spaces (IFSs), and the acknowledgment and poll frames overheads.\\
\item \textit{Admission Control}

The ACU manages the \textit{TSs} admission while maintaining the QoS of the already admitted ones. When a new \textit{TS} demands an admission, the ACU First obtains a new $SI$ as shown in the previous step and computes number of MSDUs arrived at the new $SI$ using Equation~\eqref{eq:n}. Next, it calculates the $TXOP_{i}$ for the particular \textit{TS} using Equation~\eqref{eq:txop}. Finally, ACU admits only the \textit{TS} if the inequality in Equation \eqref{eq:ACU} is satisfied.
\begin{equation}
\label{eq:ACU}
\frac{TXOP_{ n + 1}}{SI} + \sum_{i = 1}^{n} \frac{TXOP_{i}}{SI} \leq \frac{\Delta T - T_{CP}}{\Delta T}
\end{equation}
where $n$ is the number of currently admitted \textit{TSs}, so that ($n + 1$) represents the incoming \textit{TS}, $\Delta T$ is the time interval between two consecutive TBTTs periods (i.e. $\Delta T= T_{i+1}^{TBTT}-T_{i}^{TBTT}$), beacon interval and $T_{CP}$ is the duration reserved for EDCA. The HC sends an acceptance message (ADDTS-Response) to the requested QSTA if the condition in Equation~\eqref{eq:ACU} is true or send a rejection message otherwise. The accepted \textit{TS} will be added to the polling list of the HC.
\end{enumerate}

\subsection{Variable Bit Rate MPEG--4 Video Traffic}
\label{sec:VBRTraffic}

Variable Bit Rate video can be classified in terms of variability of the traffic into two types: variable in packet size such as MPEG--4; and variable on the generation interval such as H.263. As this research is aimed towards enhancing the TXOP allocation we have chosen MPEG--4 video.

MPEG--4 is an efficient video encoding covering a wide domain of bit rate coding ranging from low-bit-rate for wireless transmission up to higher quality beyond high-definition television (HDTV) \cite{Fitzek2001}. For this reason, MPEG--4 video coding has become from among the prominent videos in the internet nowadays. This variability in the compression level is adequate to transmit the video packets over the limited wireless network resources such as low bit rate. Table~\ref{tab:traceFragHigh} displays excerpt of video trace file of Jurassic Park 1 movie \cite{Fitzek2000} encoded using MPEG--4 at high quality. We refer the reader to \cite{koenen1999,Koenen2002,soares1998} for more details about MPEG--4 videos.
\begin{table}
\centering
\caption {A fragment of Jurassic Park 1 Trace File Encoded Using MPEG--4 at High Bit Rate}
\begin{tabular}{cccc}
\hline
Frame & Frame & Frame & Frame \\
sequence & type & period (ms)& size (byte) \\
\\ \hline
482 & P & 19320 & 946 \\
483 & B & 19240 & 749 \\
484 & B & 19280 & 748 \\
485 & P & 19440 & 1230 \\
486 & B & 19360 & 674 \\
487 & B & 19400 & 685 \\
488 & P & 19560 & 1208 \\
489 & B & 19480 & 815 \\
490 & B & 19520 & 804 \\
491 & I & 19680 & 3159 \\
492 & B & 19600 & 707 \\
493 & B & 19640 & 645 \\ \hline
\end{tabular}
\label{tab:traceFragHigh}
\end{table}
Basically, the weakness of HCCA in supporting VBR traffic is because of the lack of information about the abrupt changes of the traffic profile during the time, more particular the traffic burstiness issue.

\subsection{Transmission of MPEG--4 video in HCCA}
Although HCCA guarantees a QoS for video traffic based on the required TSPEC parameters, there is a probability to have frames smaller than the mean negotiated MSDU size. Consequently, a larger TXOP than needed will be assigned to a QSTA causing wasting in wireless channel time and remarkable increase in the end-to-end delay. Figure~\ref{dig:HCCAvsATXOP} (upper part) illustrates the effect of assigning TXOPs for VBR traffics based on the mean TSPEC parameters on increasing the packet delay and on the poor wireless channel utilization. Suppose there are three stations sending uplink video traffics to QAP. HC will accordingly assign $TXOP_{1}$, $TXOP_{2}$ and $TXOP_{3}$ to $QSTA_{1}$, $QSTA_{2}$ and $QSTA_{3}$ respectively. Assume that in any $SI$ some or all the frames sent is considerably smaller than the negotiated MSDU for the \textit{TS}, in this case the QSTA will only utilize a portion of the scheduled TXOP for sending its data as explained in the $SI_ {i} $ and $SI_{i + 1}$. The next scheduled TXOP will initiate according to its scheduled time, regardless the actual exploited time in the previous TXOP causing increases in the delay and wasting the channel time as well. This issue may be noticeably severe when the number of QSTA with VBR traffics increase. Moreover, in the transmission of the pre-recorded video, the traffic behavior is known prior to the traffic setup \cite{Fu2003}. These observations motivate us to present an enhancement to HCCA scheduler in which HC exploits information sent by QSTA about the changes in the traffic profile so as to accurately assign TXOP to QSTA and advance the consecutive TXOPs to minimize the delay and add the residual wireless channel time to EDCA period. The proposed scheduler is presented in details in the next section.

Several approaches such as \cite{Lee2009,Jansang2011,Cecchetti2012,cecchettielAL2012,ruscelli2013,Lee2013} have been presented in the literature attempting to remedy the deficiency of the HCCA reference scheduler in supporting QoS for VBR traffics. However, these enhancements are still not sufficient to cope with the fast fluctuating nature of highly compressed video applications since the QSTAs are scheduled according to an estimation about the uplink \textit{TSs} characteristic which may be far from the real traffics. Although, some other approaches such as \cite{Zhang2013,Ali2013,Huang2010,He2011,Chou2011} have been focused on enhancing the polling overhead of HCCA scheduler, there was a lack of consideration for transmitting pre-recorded streams in WLANs. 

More particularly, with highly variable bit rate applications, it is likely one of two cases happen at some certain SIs. One case is when the application data rate goes up in which the allocated time is not sufficient to empty the transmission queue of the uplink traffics at the QSTAs which in turn results in increasing the end-to-end delay. There are two possible solutions to remedy this problem \cite{cecchettielAL2012,ruscelli2014}. One is by maximizing the TXOP duration more than that allocated with regards to the average TXOP of the traffic. However, it will much degrade the bandwidth utilization when the data rate drops down. The other one is by applying a bandwidth reclaiming scheme \cite{Cecchetti2012,lo2007,rashid2008,ruscelli2011}. The other case is when data rate drops below its average value declared in the TSPEC of the TS. In this case, the assigned TXOP is not totally exhausted. This issue is addressed in this paper.

\subsection{Adaptive TXOP Scheduling Algorithm}
\label{sec:ATXOP}
\begin{figure*}
\centering
\includegraphics[scale=.95]{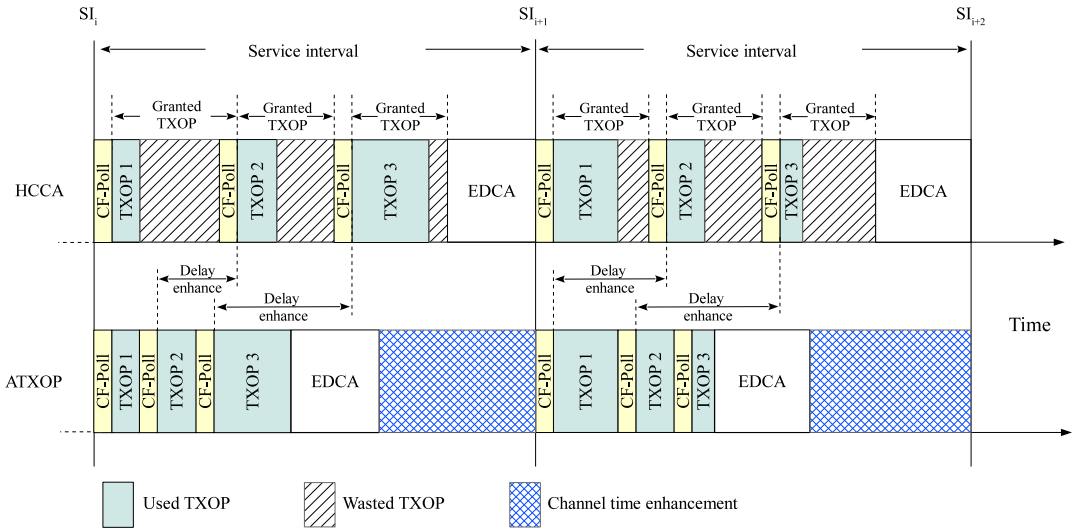}
\caption{Adaptive TXOP assignment algorithm}
\label{dig:HCCAvsATXOP}
\end{figure*}

Adaptive Transmission Opportunity (ATXOP) scheduler \cite{Almaqri2013} is a feedback-based technique which reschedules TS of each QSTA in an SI based on piggybacked information, transmitted to HC, about the next frame length. This algorithm gives an actual TXOP needed by stations and ensures that the end-to-end delay is minimized without jeopardizing the channel bandwidth. An example of the ATXOP algorithm process compared to HCCA scheduler is depicted in Figure~\ref{dig:HCCAvsATXOP}. More particularly, the QSTAs are scheduled as in Equation~\eqref{eq:txop} except that the Mean Size of MSDU ($L_{i}$) is the actual frame size reported in the previous received packet instead of using the mean value negotiated in the TS setup phase.
\section{Adaptive Multi-polling TXOP Scheduling Algorithm}
\label{sec:ATAV}
\begin{figure*}
\centering
\includegraphics[width=\linewidth]{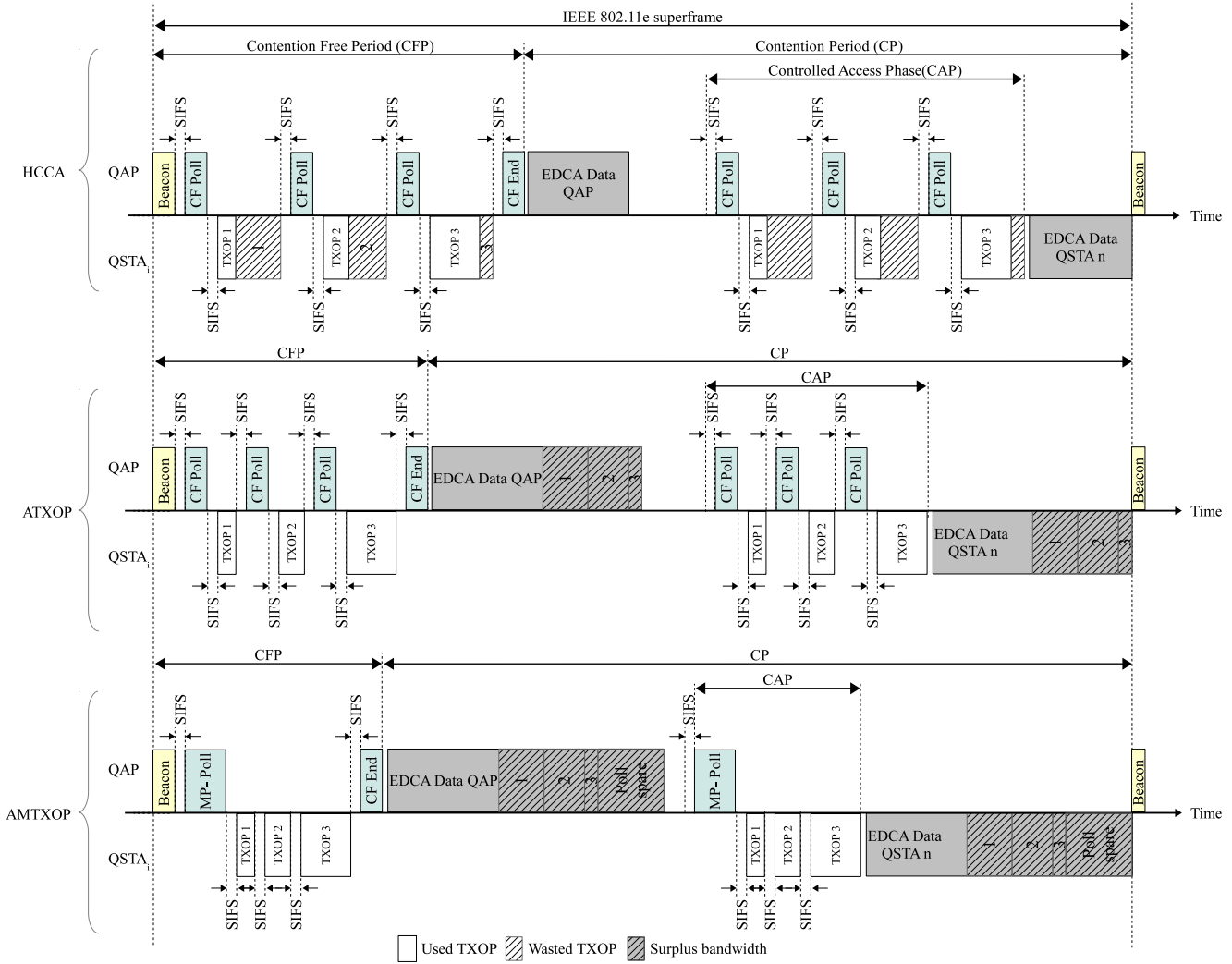}
\caption{Dynamic TXOP assignment algorithm.}
\label{dig:SinglePollvsMultiPolling}
\end{figure*}

To leverage the advantage of the global scheduler and minimize the control messages overhead, we propose a multi-polling scheme called Adaptive Multi-polling TXOP scheduling algorithm (AMTXOP) which enhances the previously investigated algorithm in \cite{Almaqri2013}. The benefits of the result scheduler over the previous scheme are two fold, on one hand, it further reduces the delay exposed by transmitting poll messages frame for all the stations that precede the current station. On the other hands, it enhances the channel utilization by freeing some space reserved for the single poll messages which replaced by only one multiple poll message.
The delay experienced by the proceeding unused (wasted) TXOPs is minimized using this algorithm as illustrated in Figure~\ref{dig:SinglePollvsMultiPolling}. The scheduling operations at both QSTA and QAP are described below.

\subsection{Multi-Polling frame structure}
\label{sec:MPFS}
\begin{figure}
\centering
\includegraphics[width=\linewidth]{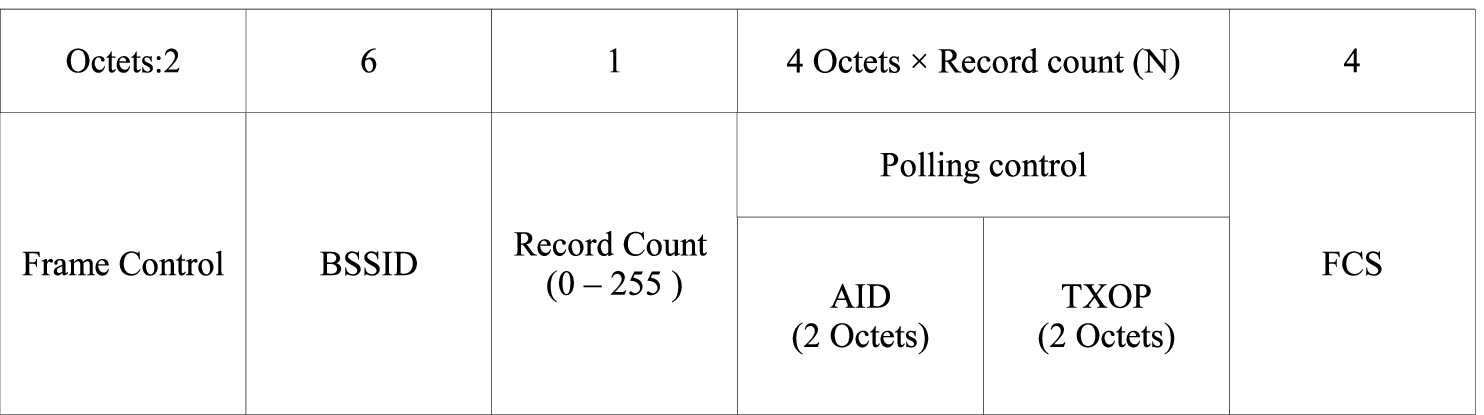}
\caption{Multi-polling frame format}
\label{dig:MultiPollFrame}
\end{figure}

At the beginning of each CAP, the HC shall have collected information about the next frame size of all admitted TSs in the polling list. Accordingly, the HC compose one multi-polling frame to be broadcast to all QSTAs. The format of the multi-polling frame is shown in Figure~\ref{dig:MultiPollFrame}.
The record number field indicates the number of the QSTAs to be polled in this service interval. The polling control includes: 1) associate identifier (AID) subfield which differentiate QSTAs in the BSS, 2) the TXOP subfield which is the TXOP duration for QSTA determined using corresponding AID. Other fields are described in details in the IEEE 802.11e standard \cite{IEEEStandard2012}.
It is worth noting that the length of the multi-polling frame is based on the number of the QSTAs admitted in the current SI. One can notice that the multi-polling frame proposed here is 1 byte smaller than that in \cite{ByungSeoKim2005} due to eliminating the rate subfield as the rate of the TS is declared in TSPEC request frame. Suppose that 802.11g wireless channel with basic physical rate of 2 is used, the transmission time of multiple single poll frames and one multi-polling frame will be as depicted in Table \ref{tab:singlevsMultipoll}.
It is obvious that the time required to transmit one multi-polling frame is for $N$ STAs is much smaller than that of multiple single frames. The multi-polling transmission time of $N$ QSTAs is computed as time to transmit $13+(N\times4)$ bytes while it is as much as transmitting $Poll_{size} \times N$ bytes in the case of single poll frames. Moreover, the reduction in poll overhead can be implied from the gain ratio.

\begin{table*}
\centering
\caption {Comparison between the transmission overhead of multiple single poll frames and one multi-polling frame measured in unit of $\mu$s.}

\begin{tabular}{llllllllll}
\hline
Number of stations ($N$) & 1 & 2 & 3 & 4 & 5 & 6 & 7 & 8 & 9 \\\hline
Single poll frame$\times N$ & 264 & 528 & 792 & 1056 & 1320 & 1584 & 1848 & 2112 & 2376 \\
Multi-polling frame & 284 & 300 & 316 & 332 & 348 & 364 & 380 & 396 & 412 \\
Gain ratio & $\simeq$ 0 & 0.43& 0.60& 0.69& 0.74 & 0.77& 0.79 & 0.81 & 0.83 \\ \hline
\end{tabular}
\label{tab:singlevsMultipoll}
\end{table*}

\subsection{Actions at the station}
At the transmission of the packet of uplink traffic, the QSTA obtain information about the next MSDU frame size from the application layer via cross-layering. This information is carried in the QS subfield of 11e MAC header introduced in IEEE 802.11 standard \cite{IEEEStandard2012}. This information is exploited by the HC for scheduling purpose, which is discussed in the following subsection.

Bearing in mind that the AMTXOP does not modify the polling list of the HCCA, the order of the polling records will represent the order the QSTA scheduled in. Assume without loss of generality that the polling records are in an ascending order based on their AID, hence, upon the reception of multi-poll frame, the $ QSTA_{i} $ computes the $backoff_{i}$ that shall wait until it can transmit its data which can be obtained from the summation of $TXOP{j}$ durations of all preceding stations, where $j \in [1..i-1]$. Then it initiates a backoff timer waiting for all QSTA ahead to finish their transmission. At the expiration of the backoff timer, it commences its transmission for a duration equals to $TXOP_{i}$. The algorithm at QSTA is reported in Algorithm \ref{algo01}.

\begin{algorithm}
\caption{ Dynamic TXOP Assignment algorithm at $QSTA_{i}$ }\label{algo01}

AT THE EVENT OF RECEIVING A MULTI-POLL FRAME

\uIf{AID of $station_{i}$ is in the polling control}{
$backoff_{i} \leftarrow + \sum\limits_{j=1}^{i-1} TXOP_{j}$
}
Start $backoff_{i}$ countdown

\uIf{$backoff_{i}$ is elapsed}{
Transmit data for duration equals to $TXOP_{i}$
}
\Else{
Wait for single poll message
}

\end{algorithm}\DecMargin{1em}

\subsection{Actions at the access point}
Upon the reception of the first frame from a $QSTA_{i}$ at HC, which includes the MSDU size of the next frame, the HC will start maintain this information in $stations$ list until the next CAP begins. At that time, the TXOP duration will be computed as expressed in Equation~\eqref{eq:atav}. 
\begin{equation}
TXOP_{i}=\frac{Size_{i}}{R_{i}}+ O
\label{eq:atav}
\end{equation}
In the case of having more than one TS at $QSTA_{i}$, the $Size_{i}$ will represent a total summation of all next frames lengths in bytes. When no data packet is received in the current $SI$ due to packet loss, the $TXOP_{i}$ of $QSTA_{i}$ will be computed as expressed in Equation~\eqref{eq:txop}. It is worth noting that at the first CAP of any \textit{TS}, the TXOP is calculated based on Equation~\eqref{eq:txop} because no information about the next packet size has been reported yet.
\begin{algorithm}
\caption{ Dynamic TXOP Assignment algorithm at Access Point}\label{algo02}
\textbf{INPUT:}

$Stations$, a list of $N$ stations in the polling list maintained at the HC\;
$Sizes$, a list of next packet sizes for each $station_{i}$ in the $Stations$ list where $i = 1 .. N$ 
AT THE EVENT OF RECEIVING A DATA PACKET FROM $station_{i}$

Save the packet size of $station_{i}$ in $Sizes$

\For{each CAP}{
\While{$Stations$ is not null}{
\uIf{no data packet received from $station_{i}$}{
$L_{i} \leftarrow MSDU_{i}$

$TXOP_{i}=max\left (\frac{N_{i} \times L_{i}}{R_{i}}, \frac{M}{R_{i}} \right) + O$ 
}
\Else{
$TXOP_{i}=\frac{Size_{i}}{R_{i}}+ O$
}
}
Construct one multi-polling frame for all QSTAs

$t_{end} \leftarrow t_{curr}+ T_{poll} + \sum\limits_{i=1}^{N} TXOP_{i}$

Broadcast a multi-polling frame
}
\end{algorithm}\DecMargin{1em}
Instead of sending a single poll for each station, the HC compose and broadcast one multi-polling frame which includes $AID$ and computed $TXOP_{i}$.

\section{Performance Evaluation}
\label{sec:evaluation}
In order to evaluate the performance of the AMTXOP scheduler, we have used a network simulation tool. The simulation environment setup, and video traffic used as uplink traffics is described in details in this section. The performance of our scheduler is compared against the HCCA. The results of end-to-end delay and throughput are also discussed.
\subsection{Simulation Setup}
\label{sec:simSet}
The AMTXOP scheduler has been implemented in the well-known network simulator (\textit{ns-2}) \cite{NS2} version 2.27. The HCCA implementation framework \textit{ns2hcca} \cite{cicconetti2005} has been patched to provide the controlled access mode of IEEE 802.11e functions, HCCA. The \textit{ns-2} Traffic Trace \cite{NSBook2012} agent, is used for video stream generation.

\begin{figure}
\centering
\includegraphics[scale=0.30]{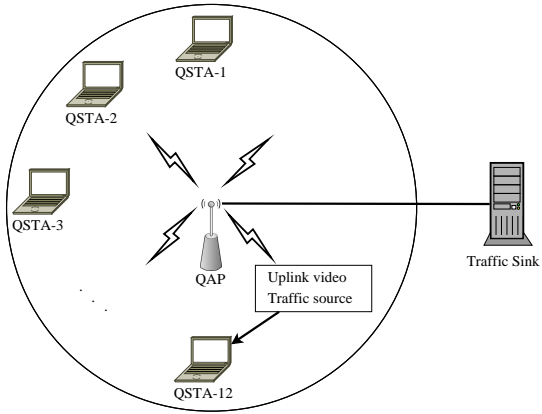}
\caption{Network topology.}
\label{fig:topology}
\end{figure}

A star topology has been used for constructing the simulation scenario to form an infrastructure network with one QAP surrounded by varying number of the QSTAs ranging from 1 to 12. All QSTAs were distributed uniformly around the QAP with a radius of 30 feet as shown in Figure~\ref{fig:topology}. The stations were placed within the QAP coverage area, in the same basic service set BSS, and the wireless channel is assumed to be ideal. Since we focus on HCCA performance measurement, all the stations operate only on the contention-free mode by setting $T_{CP}$ in Equation~\eqref{eq:ACU} to zero. The QSTAs communicate directly without a hidden node problem. RTS/CTS mechanism, MAC level fragmentation, and multirate support were disabled.
QAP is the sink receiver, while each QSTA is a video source because only one flow per station is supported by \textit{ns2hcca} patch. Therefore, for simulating concurrent video streams multiple stations are added each with one flow. In order to leave an ample time for initialization, stations start their transmission after 20 (sec) from the start of the simulation time and last until the simulation end. Wireless channel assumed to be an error-free, and no admission control used for the sake of investigating the maximum scheduling capability of each examined algorithm under heavy traffic conditions. Simulation parameters are summarized in Table~\ref{tab:SimPars}.
\begin{table}
\caption {Simulation Parameters}
\centering

\begin{tabular}{ll}
\hline
Parameter & Value \\ \hline
Simulation time & 500 (sec) \\
Physical layer & IEEE 802.11g \\
MAC layer & IEEE 802.11e \\
SIFS & $10\mu s$ \\
PIFS &  $30\mu s$ \\
Slot time & $9\mu s$ \\
Physical preamble length& 12 (bytes) \\
PLCP header length & 3 (bytes) \\
PLCP data rate & 1 (Mbps) \\
MAC header size & 36 (bytes) \\
Data rate & 54 (Mbps) \\
Basic physical rate & 1 (Mbps) \\
Propagation delay & $2\mu s$\\ \hline
\end{tabular}
\label{tab:SimPars}
\end{table}
For evaluating the performance of our scheduler against the reference scheduler of HCCA, Jurassic Park 1 video sequence trace encoded using MPEG--4 was chosen from a publicly available library for video traces \cite{Fitzek2001}. We tested the AMTXOP scheduler on \textit{Jurassic Park 1} and \textit{Formula 1 } trace files which can be classified into movie and sport, respectively, which show different variability level. Table~\ref{tab:traceStats} demonstrates some statistics of the examined traces. The selected video is encoded using different Compression ratio, which results in varying quality. 
\begin{table}
\centering

\caption {Frame Statistics of MPEG--4 Video Trace File}
\begin{tabular}{ l|lllll }
\hline
Video & Parameter & \multicolumn{2}{c}{Quality} \\
\cline{3-4}
& & Low & High \\ \hline
\multirow{6}{*}{\rotatebox[]{90}{Jurassic Park 1}}& Comp. ratio & 49.46 & 9.92 \\
& Mean size (byte) & 7.7e+02 & 3.8e+03 \\
& CoV of frame size & 1.39 & 0.59 \\
& Mean bit rate (bit/sec) & 1.5e+05 & 7.7e+05 \\
& Peak bit rate (bit/sec) & 1.6e+06 & 3.3e+06 \\
& Peak/Mean of bit rate & 10.61 & 4.37 \\ \hline
\multirow{6}{*}{\rotatebox[]{90}{Formula 1}} & Comp. ratio & 43.51 & 9.92 \\
& Mean size (byte) & 8.7e+02 & 4.2e+03 \\
& CoV of frame size & 1.12 & 0.42 \\
& Mean bit rate (bit/sec) & 1.7e+05 & 8.4e+05 \\
& Peak bit rate (bit/sec) & 1.4e+06 & 2.9e+06 \\
& Peak/Mean of bit rate & 8.05 & 3.45 \\ \hline
\end{tabular} 
\label{tab:traceStats} 
\end{table}
In this paper, we have tested the algorithms with low and high-quality video trace. It is worth noting that the variability of the selected videos is measured by the Coefficient of Variation (CoV), which is the standard deviation of the frame size divided by the average frame size. TSPEC parameters used for each video traffic are shown in Table~\ref{tab:VideoParas} with regards to video QoS requirements.
\begin{table}
\centering

\caption { Video Traffic Parameters }
\begin{tabular}{l|lllc}
\hline
Video 	& Parameter	& Unit	&\multicolumn{2}{c}{Quality} \\
\cline{4-5}
		& 			& 		& Low & High 				 \\ \hline
\multirow{6}{*}{\rotatebox[]{90}{Jurassic Park 1}}	& $L$		& bytes		& 7.7e+02	& 3.8e+03	\\
													& $M$		& bytes		& 8154		& 16745		\\
													& $\rho$	& bit/sec	& 1.5e+05	& 7.7e+05	\\
													& $D$		& sec		& 0.08		& 0.08 		\\
													& $R$		& Mbps		& 11		& 11		\\
													& $MSI$		& sec		& 0.04		& 0.04		\\ \hline
\multirow{6}{*}{\rotatebox[]{90}{Formula 1}} 		& $L$		& bytes		& 8.7e+02 	& 4.2e+03 	\\
													& $M$		& bytes		& 7032 		& 14431 	\\
													& $\rho$	& bit/sec 	& 1.7e+05 	& 8.4e+05 	\\
													& $D$		& sec		& 0.08 		& 0.08 		\\
													& $R$		& Mbps 		& 11 		& 11 		\\
													& $MSI$		& sec		& 0.04 		& 0.04 		\\ \hline
\end{tabular}
\label{tab:VideoParas}
\end{table}
\subsection{Results and Discussion}
Simulations have been carried out to exhibit the performance of the examined algorithms, namely HCCA, an Adaptive TXOP assignment which we refer to as ATXOP onward and the proposed algorithm using a different variability level of the same videos. Since the main objective is to achieve superior QoS support by accurately granting TXOP to the station to fit its need, packet end-to-end delay of the uplink traffics has been measured which considered as one of the significant metrics to evaluate a QoS support of video streams. To validate the behavior of the examined algorithms, the measurements are done for an increasing number of \textit{TSs}. The system throughput was also investigated to verify that the improvement in the delay is achieved without jeopardizing the channel bandwidth.

\begin{figure}
\centering
\subfigure[Low-quality.]{
\label{fig:AllocatedTXOPLow1}
\includegraphics[scale=0.31, angle=-90]{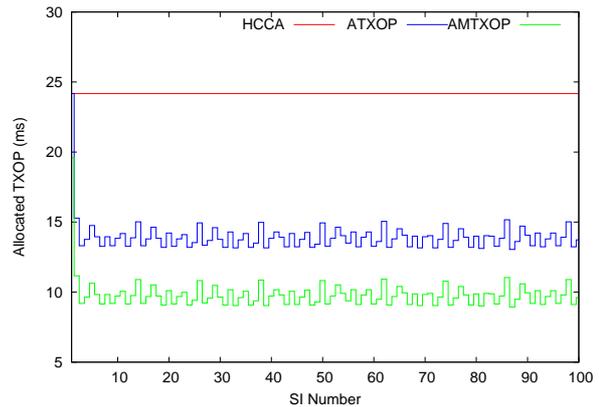}
}
\subfigure[High-quality.]{
\label{fig:AllocatedTXOPHigh1}
\includegraphics[scale=0.31, angle=-90]{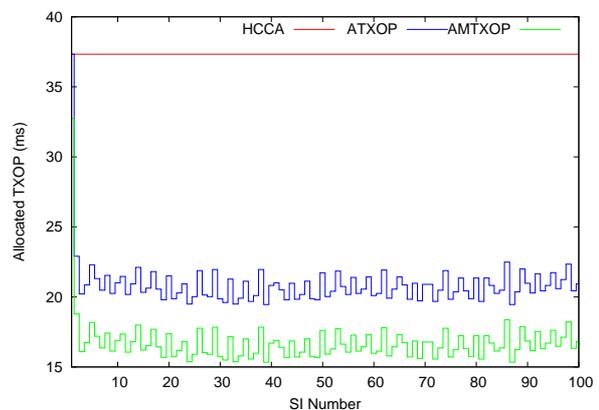}
}
\caption{TXOP allocation of Formula 1 video}
\label{fig:allocTXOP}
\end{figure}

The behavior of the examined algorithms in terms of allocating TXOP in each SI is illustrated in Figure~\ref{fig:allocTXOP} for the Formula 1 video sequence. The allocated TXOP for one flow is shown against a number of SIs for a duration of 30 seconds. The results reveal the fact of assigning fixed TXOP in HCCA for all SIs of the flow with accordance to Equation~\eqref{eq:txop}. In this case, the HCCA computes TXOP duration based on the maximum MSDU size of the flow, namely 7032 bytes and 14431 bytes for low and high-quality video respectively. Nevertheless, the both ATXOP and the proposed scheduler alike adaptively allocate a TXOP for each SI based on the actual frame size obtained from the feedback information which show that in some SIs the allocated TXOP duration in HCCA is much higher than the actual need of the flow which considered as over-allocation cases. It is obvious that the TXOP duration given in \ref{fig:AllocatedTXOPHigh1} is higher than that in \ref{fig:AllocatedTXOPLow1} as the mean bit rate of high-quality encoded Formula 1 is considerably higher than that in low-quality video sequence, refer to Table \ref{tab:traceStats}.

\subsubsection{End-to-End Delay Analysis}
\label{sec:e2edly}
\begin{figure}
\centering
\subfigure[Low-quality Jurassic Park 1.]{
\label{fig:e2eLowDly1}
\includegraphics[scale=0.31, angle=-90]{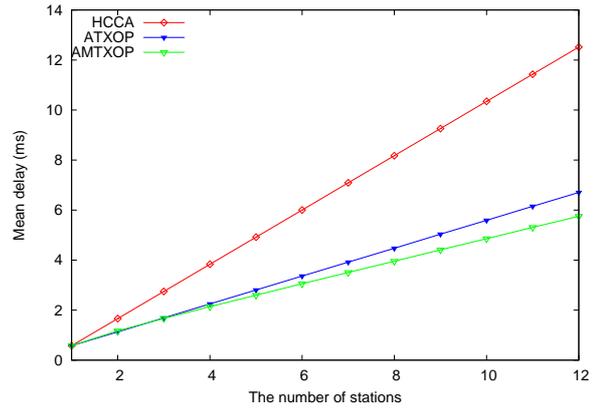}
}
\subfigure[High-quality Jurassic Park 1.]{
\label{fig:e2eHighDly1}
\includegraphics[scale=0.31, angle=-90]{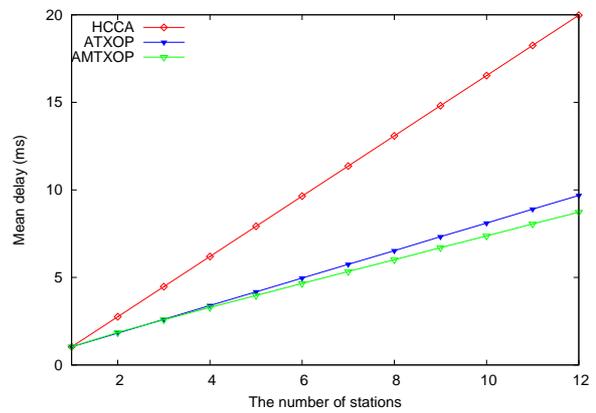}
}
\subfigure[Low-quality Formula 1.]{
\label{fig:e2eLowDly2}
\includegraphics[scale=0.31, angle=-90]{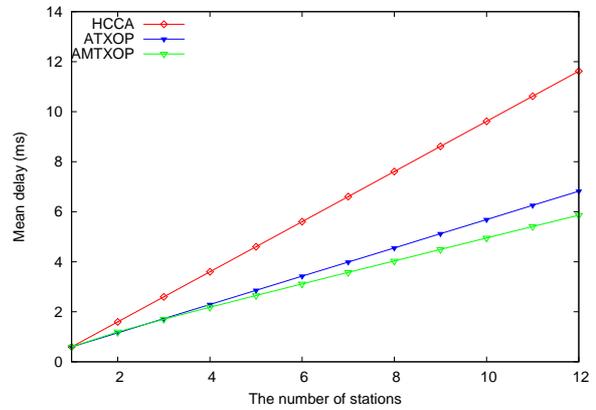}
}
\subfigure[High-quality Formula 1.]{
\label{fig:e2eHighDly2}
\includegraphics[scale=0.31, angle=-90]{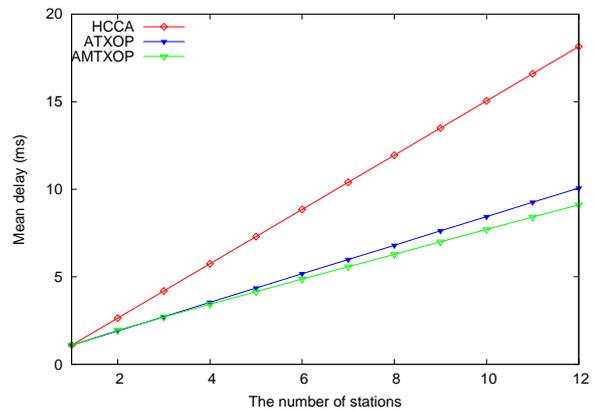}
}
\caption{Mean end-to-end delay as a function of the number of stations.}
\end{figure}

The end-to-end delay is defined as the time elapsed from the generation of the packet at the source QSTA application layer until it has been received at the QAP and is expressed in Equation~\eqref{eq:meanDelay}.
\begin{eqnarray}
\label{eq:meanDelay}
e2eDelay = \frac { \sum_{i = 1}^{N} ( R_{i} - G_{i}) } {N},
\end{eqnarray}
where $G_{i}$ is the generation time of packet $i$ at the source QSTA, $R_{i}$ is the receiving time of the particular packet at the MAC layer of the QAP, and $N$ is the total number of packets for all flows in the system. The end-to-end delay has been measured for the three video types to study the efficiency of both HCCA and our algorithms with different traffic variability. Figure~\ref{fig:e2eLowDly1}, \ref{fig:e2eHighDly1} and \ref{fig:e2eLowDly2}, \ref{fig:e2eHighDly2} depict the delay experienced by data packets for the low, medium and high-quality video, respectively. One can notice that the end-to-end delay boosts with the increase of the packet size, the highest quality exhibits higher end-to-end delay and vice versa. The increase of the delay in higher quality videos can be justified by the large amount of the allocated to each \textit{TS}, as in Equation~\eqref{eq:txop}, which leads to maximize the wasted TXOPs that keep the subsequent \textit{TSs} awaiting in their transmission queue longer time. It is obvious that the proposed scheduler achieved delay enhancement over both HCCA and ATXOP for the examined videos due to the polling overhead reduction in the system. The proposed algorithm achieved delay enhancement up to 12\% over the Adaptive TXOP algorithm \cite{Almaqri2013} and about 59\% over HCCA scheduler. The reason of achieving better improvement in Jurassic Park 1 is the higher packet size comparable to that in Formula 1 thereby the granted TXOP obtained in HCCA is far from the needed TXOP which in turn causes higher packet delay. Furthermore, the delay improvement in our scheduler is justified by the accurate calculation of the TXOP. Unlike the HCCA scheduler that only relies on the mean traffic characteristic which is not reflecting the actual traffic behavior.

\subsubsection{Throughput Analysis}
\begin{figure}
\centering
\subfigure[High-quality Jurassic Park 1.]{
\label{fig:thrpHigh1}
\includegraphics[scale=0.31, angle=-90]{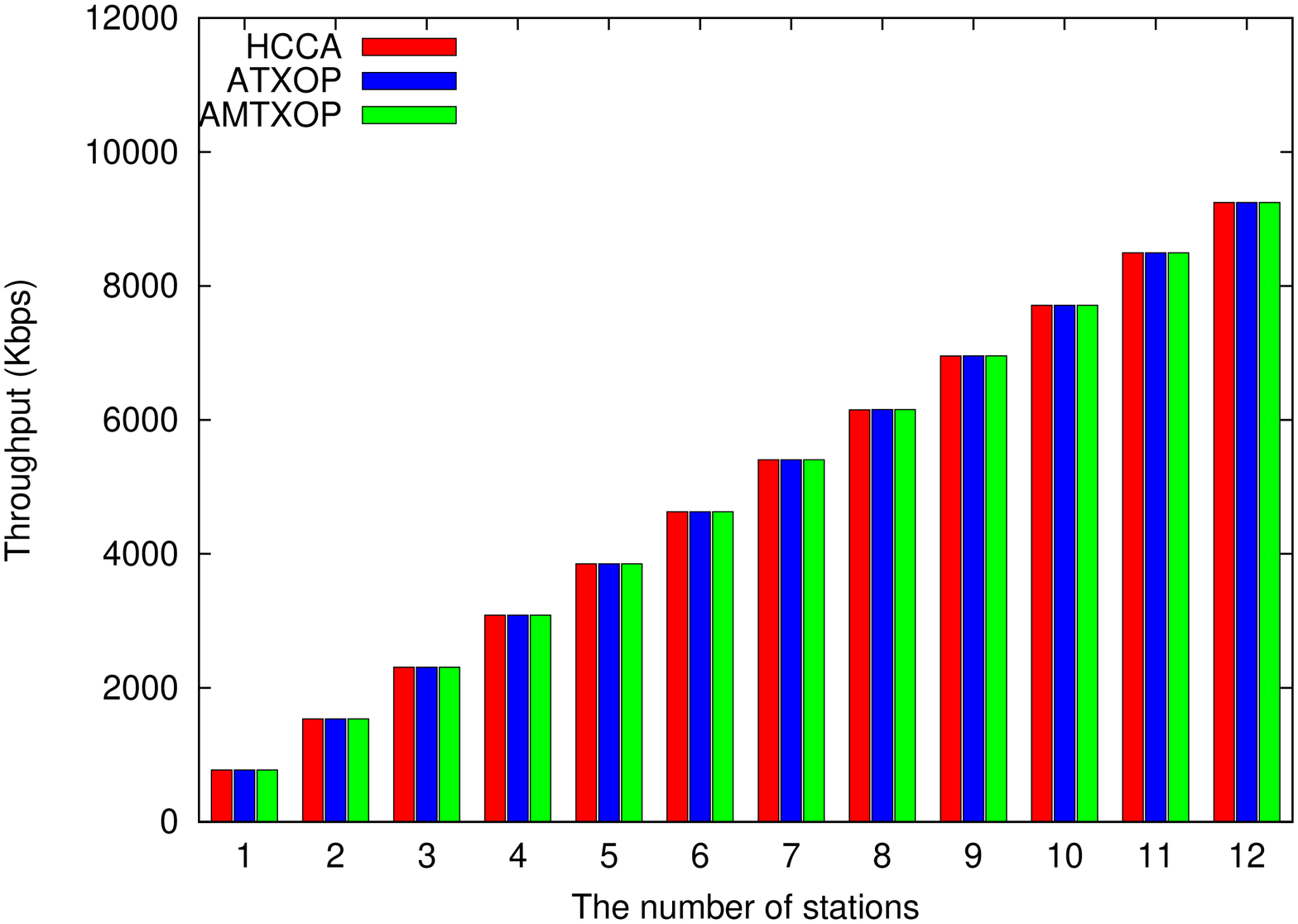}
}

\caption{Aggregate throughput as a function of the number of stations.}
\end{figure}

The aggregate throughput of the examined algorithms has been investigated against the number of stations. This is to verify that our scheduler is efficient in supporting QoS for VBR traffics which maintaining the utilization of the channel bandwidth. The aggregate throughput is calculated using Equation~\eqref{eq:throughput}.
\begin{eqnarray}
\label{eq:throughput}
AggregateThrp = \frac { \sum_{i = 1}^{N} ( Size_{i}) } {time},
\end{eqnarray}
where $Size_{i}$ is the received packet size at the QAP, $time$ is the simulation time and $N$ is the total number of the received packets at QAP during the simulation time. Figure \ref{fig:thrpHigh1} depicts the aggregate throughput with increasing the network load for the high-quality Jurassic Park 1 videos. The result shows that the throughput is the same as that achieved by the HCCA scheduling algorithm. This implies that the proposed approach enhanced the end-to-delay without jeopardizing the channel bandwidth.

\subsubsection{Aggregate TXOP Duration}
\begin{figure}
\centering
\subfigure[Low-quality Jurassic Park 1.]{
\label{fig:TXOPLow1}
\includegraphics[scale=0.31, angle=-90]{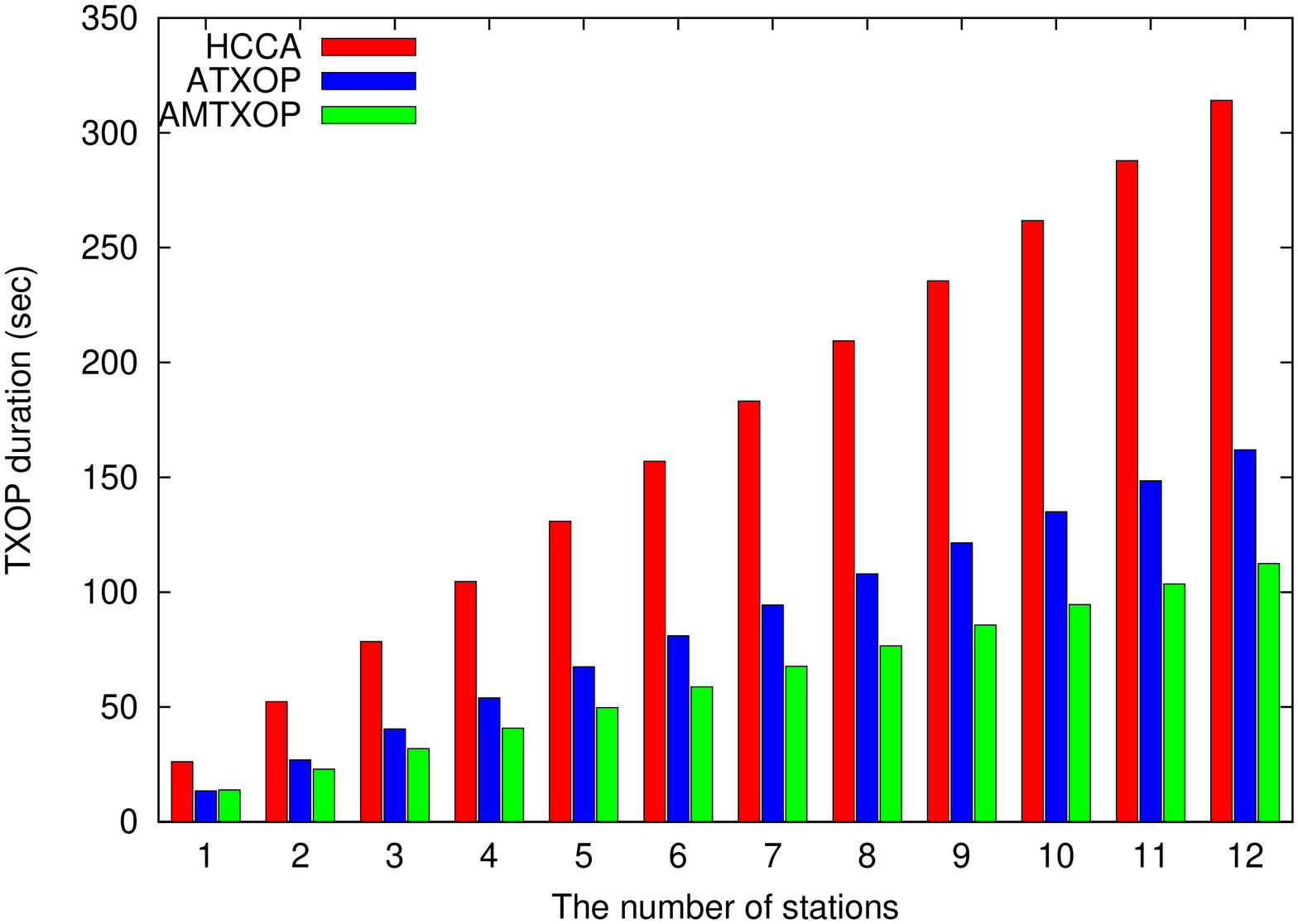}
}
\subfigure[High-quality Jurassic Park 1.]{
\label{fig:TXOPHigh1}
\includegraphics[scale=0.31, angle=-90]{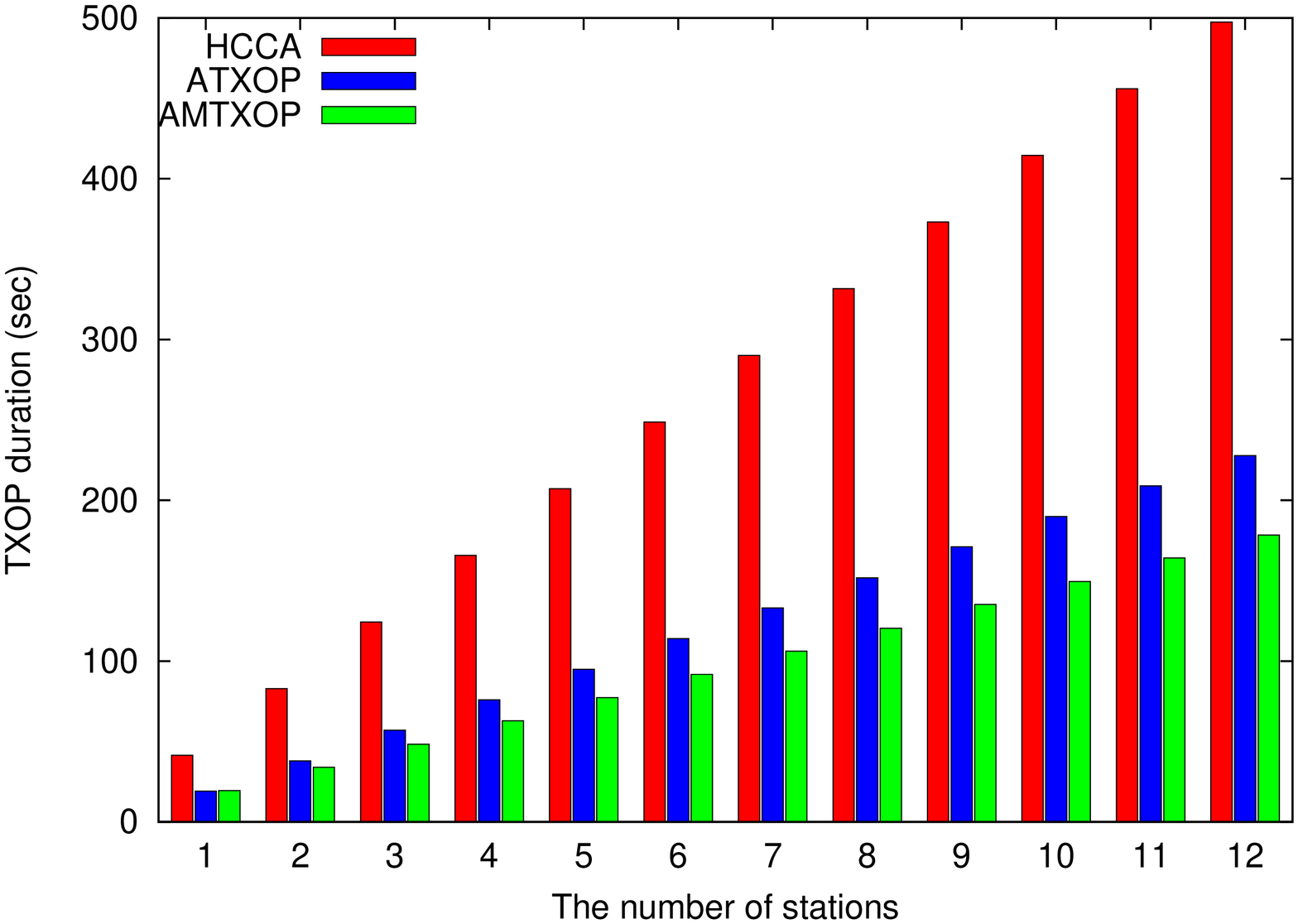}
}
\subfigure[Low-quality Formula 1.]{
\label{fig:TXOPLow2}
\includegraphics[scale=0.31, angle=-90]{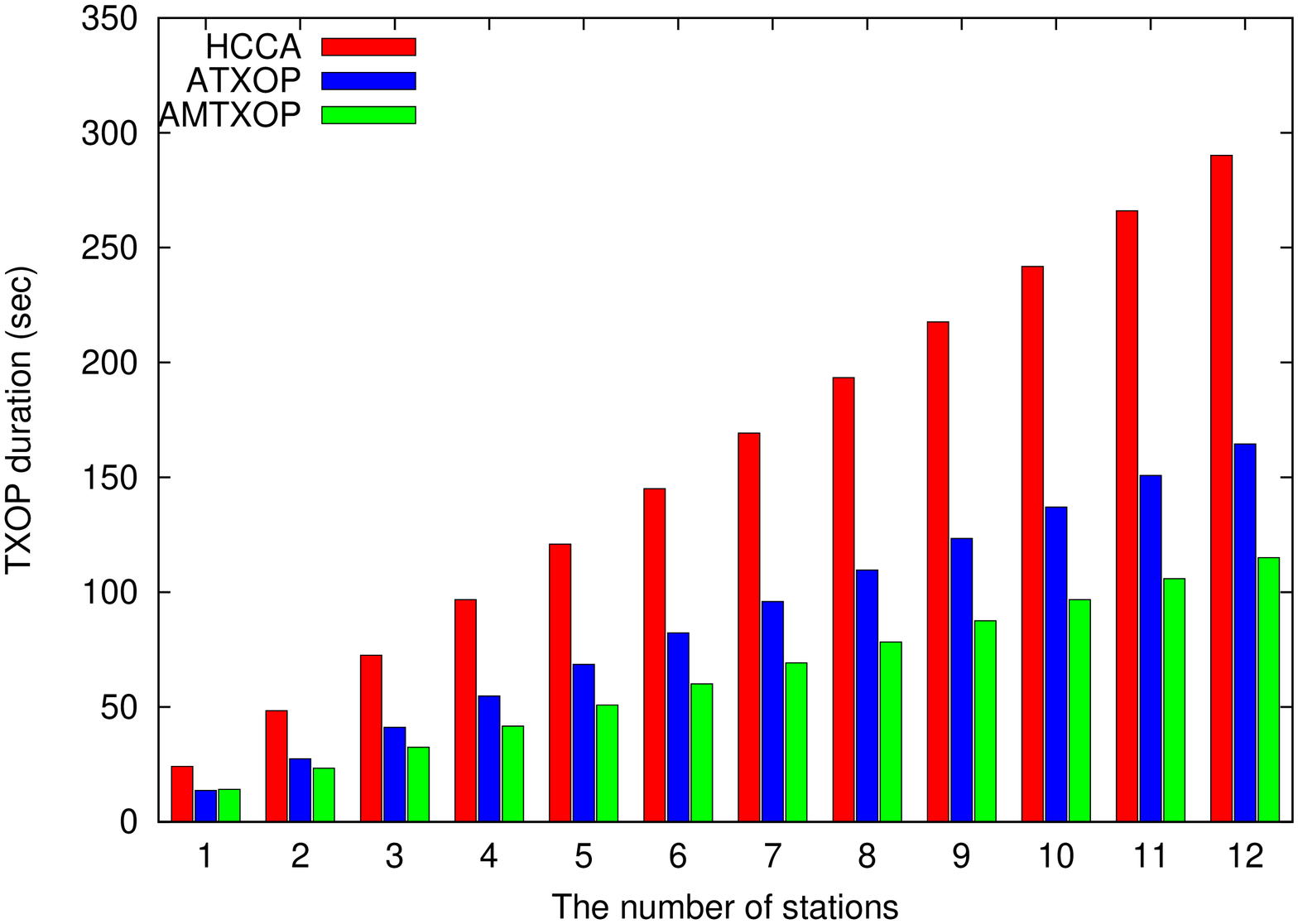}
}
\subfigure[High-quality Formula 1.]{
\label{fig:TXOPHigh2}
\includegraphics[scale=0.31, angle=-90]{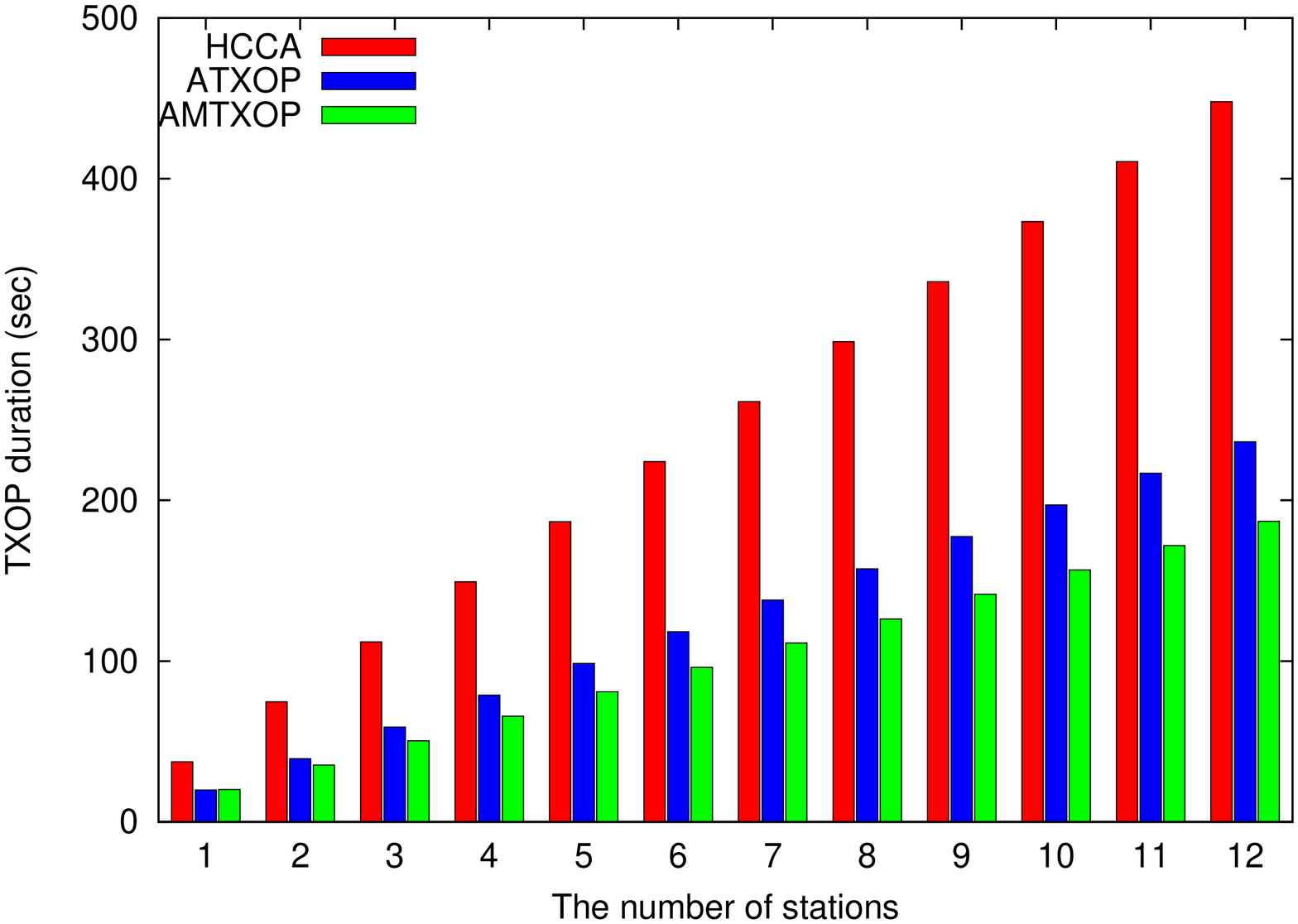}
}
\caption{Aggregate TXOP duration as a function of the number of stations.}
\label{fig:AggTXOP}
\end{figure}

To investigate the efficiency attained by the proposed scheduler in supporting pre-recorded videos, the aggregate TXOP duration is measured and can be defined as the total of TXOP duration, which is the total TXOP duration granted to all QSTAs during the simulation time in units of seconds. In Figure~\ref{fig:AggTXOP}, the aggregate TXOP is shown in the examined videos with increasing the network load. For Low-quality videos Figures~\ref{fig:TXOPLow1} and \ref{fig:TXOPLow2} demonstrate that allocating fixed TXOP for all \textit{TS} frames in HCCA might exceed the need of the traffic. In that case only a small portion of the granted TXOP is exploited resulting in what's called wasted TXOPs. On the contrary, since the proposed scheduler not only operates according to the actual information about frame size, but also further minimizes the polling overhead, the granted TXOP is considerably smaller than that in HCCA and ATXOP with ensuring the same achievable throughput. Thanks to multi-polling concept, the total TXOP duration assigned was as low as 32\% and 25\% compared to Adaptive TXOP scheduler in low and high videos, respectively. In the proposed algorithm TXOP duration is smaller than both HCCA and that in \cite{Almaqri2013} as the polling overhead in equation \eqref{eq:txop} is minimized by sending only one multi-poll frame for all QSTAs. In this case, the proposed algorithm shows better enhancement even when the number of station increase in the system.

\subsubsection{Number of admitted MSDUs}
IEEE 802.11 Physical layer standards differ from each other in terms of physical data rate supported and interframe spaces such as SIFS and PIFS. It is worth noting that the physical layer may support various range of data rate as in IEEE 802.11g while slot time, SIFS and PIFS (referred to in \cite{IEEEStandard2012} as aSlotTime, aSIFTime and PIFS) are fixed per PHY \cite{IEEEStandard2012}. It is obvious in Equation \eqref{eq:txop} that the TXOP duration of each QSTA is inversely proportional to the PHY rate $R_{i}$ and directly proportional to SIFS and PIFS. To study this effect, the examined schemes have been also tested over 802.11b. Similar simulation parameters used as listed in Table~\ref{tab:SimPars} except physical layer, Slot time, PIFS, Physical preamble length, PLCP header length and Data rate were IEEE 802.11b, 20 $\mu s$, 30 $\mu s$, 18 bytes, 6 bytes and 11Mbps respectively. Simulation results in Table \ref{tab:e2e_111b} reveal that 802.11b only supports 5 Jurassic Park 1 videos while 802.11g can support more than 12 video traffics. In 802.11b, the calculated TXOP duration for the MSDU was 0.00714 second in this case the ACU will reject all new MSDU starting from the 6th MSDU as the CAP will become 0.04284 s which exceed the SI (0.04 second). Due to the high physical rate supported in 802.11g and smaller IFs, the number of accepted MSDU is higher.
\begin{table}
\centering
\caption { End-to-end Delay in Units of Milliseconds with Different Physical Layer}
\begin{tabular}{c|p{0.7cm}p{0.7cm}p{0.7cm}|p{0.7cm}p{0.7cm}p{0.7cm}}
\hline
& \multicolumn{3}{c}{802.11b} & \multicolumn{3}{|c}{802.11g}\\
\cline{2-4} \cline{5-7}
\begin{turn}{60}QSTA\end{turn}& 
\begin{turn}{60}HCCA \end{turn} &
\begin{turn}{60}ATXOP\end{turn} &
\begin{turn}{60}AMTXOP\end{turn} &
\begin{turn}{60}HCCA \end{turn} &
\begin{turn}{60}ATXOP\end{turn} &
\begin{turn}{60}AMTXOP\end{turn} \\ \hline
1	& 1.20	& 1.20	& 1.20	& 0.58	& 0.58	& 0.58 \\
2	& 4.76	& 2.13	& 2.17	& 1.66	& 1.13	& 1.17 \\
3	& 8.32	& 3.07	& 3.03	& 2.75	& 1.69	& 1.67 \\
4	& 11.89	& 4.02	& 3.87	& 3.83	& 2.25	& 2.14 \\
5	& 15.45	& 4.95	& 4.68	& 4.92	& 2.80	& 2.60 \\
6	& - 	& - 	& - 	& 6.00	& 3.36	& 3.06 \\
7	& - 	& - 	& - 	& 7.09	& 3.92	& 3.51 \\
8	& - 	& - 	& - 	& 8.17	& 4.47	& 3.96 \\
9	& - 	& - 	& - 	& 9.26	& 5.04	& 4.41 \\
10	& - 	& - 	& - 	& 10.35	& 5.59	& 4.86 \\
11	& - 	& - 	& - 	& 11.43	& 6.15	& 5.31 \\
12	& - 	& - 	& - 	& 12.52	& 6.71	& 5.75 \\ \hline
\end{tabular}
\label{tab:e2e_111b}
\end{table}

\subsection{Analytical Study of the Proposed Schemes}
Without loss of generality, assume that each $QSTA_{i}$ in any SI has only one TS. In other words, $N_{i}$ in Equation~\eqref{eq:txop} is set to 1, $\forall i \in [1..N]$, where $N$ is number of admitted TSs. Bearing in mind that the transmission of MSDU frame in IEEE 802.11 begins with a transmission of Physical Layer Convergence Procedure (PLCP) preamble and header parts, followed by the data payload. Therefore the transmission time of MSDU frame of size L, all in bytes, is calculated as follows:
\begin{equation}
T_{MSDU}=T_{preamble} + T_{PLCP} + T_{MAC} + T_{L}, 
\label{eq:T_MSDU}
\end{equation}
where $T_{preamble}$, $T_{PLCP}$, $T_{MAC}$ and $T_{L}$ is the transmission time of PLCP preamble, PLCP header, IEEE 802.11e MAC header and data payload (bytes) respectively which can be calculated as follows:
\begin{equation}
T_{preamble}=\frac{preamble*8}{PHY_{basic}}
\label{eq:TPreamble}
\end{equation}

\begin{equation}
T_{PLCP}=\frac{PLCP*8}{PHY_{basic}}
\label{eq:TPLCP}
\end{equation}

\begin{equation}
T_{MAC}=\frac{MAC*8}{PHY_{rate}}
\label{eq:TMAC}
\end{equation}

\begin{equation}
T_{L}=\frac{L*8}{PHY_{rate}}
\label{eq:TL}
\end{equation}

The transmission time taken for ACK packet and poll message is calculated similarly as in Equations \eqref{eq:T_ACK} and \eqref{eq:T_Poll}.

\begin{equation}
T_{ACK}=T_{preamble} + T_{PLCP} + T_{MAC},
\label{eq:T_ACK}
\end{equation}

\begin{equation}
T_{poll}=T_{preamble} + T_{PLCP} + T_{MAC},
\label{eq:T_Poll}
\end{equation}

Therefore, the TXOP duration of $QSTA_{i}$ is computed as follows
\begin{equation}
TD_{i}=T_{L} + T_{poll} + T_{ACK} + 3 \times SIFS + D_{p}
\label{eq:TD}
\end{equation}
In order to ensure the accuracy of the calculation, the propagation delay, $D_{p}$, between the source and receiving channel should be considered which is normally a small fixed value \cite{IEEEStandard2012} as shown in Table \ref{tab:SimPars}.

Accordingly, analytical study of the end-to-end delay for examined schemes will be discussed in the following Subsections.

\subsubsection{Delay Analysis of HCCA}

Assuming that the transmission of the traffic for all QSTAs begins as the same time, the end-to-end delay of a $QSTA_{i}$ can be computed as the total TXOP durations of the preceding STAs in the polling table including the transmission time of MSDU of the $QSTA_{i}$. For the sake of the simplicity, we assume that the stations are scheduled in ascending order according to their index. That is, the preceding delay of $QSTA_{i}$ in an SI is computed as in the following Equation.

\begin{equation}
D_{SI}^{i}=\sum_{j=1}^{i-1} TD_{i} + T_{L}^{i} + T_{poll} + 2 \times SIFS
\label{eq:SIdlyHCCA}
\end{equation}

As known, any period of time, $T$, contains a number of consecutive SIs, $M$, which can be computed for any experiment duration $T$ as $\left \lceil T/SI \right \rceil$. Thus, the average aggregated end-to-end delay for the a period of time $T$ can be calculated as follows:

\begin{equation}
D= \frac{1}{M} \cdot \sum\limits_{k=1}^{M} \sum\limits_{i=1}^{N} D_{SI}^{i}
\label{eq:e2edlyHCCA}
\end{equation}

\subsubsection{Delay Analysis of ATXOP}

Based on the discussed deficiency of the HCCA accommodating to the fast changing in VBR traffic profile and the solution proposed in ATXOP in Section \ref{sec:ATXOP}. When $QSTA_{i}$, at any SI, exploits only portion of its allocated $TXOP_{i}$ at the traffic setup time, namely $T_{eff}^{i}$, leaving an unspent amount of $T_{u}^{i}$. Thus, the following relation can be held:
\begin{equation*}
\begin{split}
\sum_{i=1}^{N}{T}'_{i} & = T_{eff}^{1} + T_{eff}^{2} +\cdots + T_{eff}^{N}\\
& = TD_{1} - T^{1}_{u} + TD_{2} - T^{2}_{u} +\cdots +TD_{N} - T^{N}_{u} \\
& = \sum_{i=1}^{N} TD_{i} - \sum_{i=1}^{N}T^{i}_{u} \\
\end{split} 
\end{equation*}
Where  $T^{i}_{u} \geq 0$, $\sum_{i=1}^{N}{T}D_{i}$ and $\sum_{i=1}^{N}{T}'_{i}$ is the total TXOP scheduled in any SI used in HCCA and ATXOP, respectively.
It is worth noting that $TD_{i}$ is calculated from Equation~\eqref{eq:TD} which includes the poll overhead.

Thus, the delay of $QSTA_{i}$ in an SI is computed as follows,

\begin{equation}
D_{SI}^{i}=\sum_{j=1}^{i-1} (TD_{i} - T^{i}_{u}) + T_{L}^{i} + T_{poll} + 2 \times SIFS
\label{eq:SIdlyAXTOP}
\end{equation}

The worst case scenario when all QSTAs used their $TXOP_{i}$ duration that is, $ \sum_{i=1}^{N}T^{i}_{u}=0$ and consequently,
$\sum_{i=1}^{N}{T}'_{i} = \sum_{i=1}^{N}{TD}_{i}$. In this case, ATXOP performs similarly to HCCA, Thus the ATXOP ensures at least HCCA performance.

Having given different $D_{SI}^{i}$ for HCCA and ATXOP, the end-to-end delay of both schemes can be computed as in Equation \eqref{eq:e2edlyHCCA}.

\subsubsection{Delay Analysis of AMTXOP}

Assume that the transmission time of a single and multi-polling frame is $T_{poll}$ and 
$T_{mpoll}$, respectively. Thus AMTXOP calculates the total TXOP in any SI as follows:
\begin{equation}
\begin{split}
\sum_{i=1}^{N}{T}''_{i}
& = \sum_{i=1}^{N}({T}'_{i} - T_{poll}) + T_{mpoll}
\end{split}
\end{equation}

Thus, the delay of $QSTA_{i}$ in an SI is computed as follows,

\begin{align}
D_{SI}^{i}= & T_{mpoll}+\sum_{j=1}^{i-1}(TD_{i}-T^{i}_{u})- (i-1)\times T_{poll} \nonumber \\ 
& + T_{L}^{i} + SIFS
\label{eq:SIdlyAMXTOP}
\end{align}

Eventually, the delay enhancement in any SI when multi-polling scheme is integrated can be expressed as follows:

\begin{equation}
D= \frac{1}{M} \cdot \sum\limits_{k=1}^{M} \sum\limits_{i=1}^{N} D_{SI}^{i} + T_{L}^{i} +  SIFS
\label{eq:e2edlyAMTXOP}
\end{equation}

The simulation results were validated on low and high quality jurassic park 1 video sequence using the setting in Tables \ref{tab:SimPars} and \ref{tab:traceStats} except for the simulation time which set to 50 sec. The traffics start at second 20 from the beginning of the experiment run to leave an ample time for initialization process. For this reason, the number of SIs ($M$) was 750 which calculated as $30 sec*1000/40 ms$. 

\begin{figure}
\centering
\includegraphics[scale=0.65]{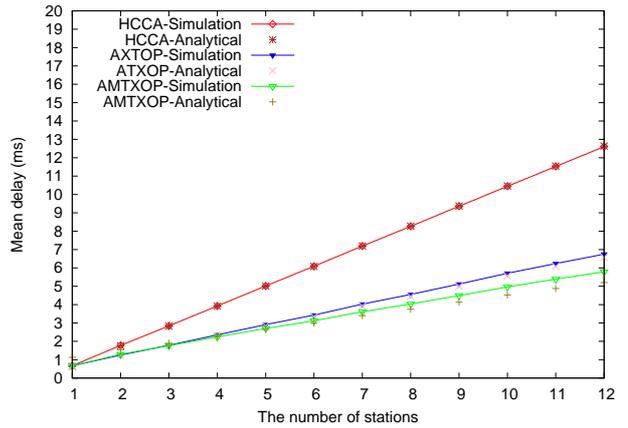}
\caption{End-to-end delay validation for Low-quality Jurassic Park 1}
\label{fig:analyticalLow}
\end{figure}

Figures \ref{fig:analyticalLow} and \ref{fig:analyticalhigh} show both simulation; and analytical results for the HCCA, AXTOP scheduling schemes were computed using Equation \eqref{eq:e2edlyHCCA} and AMTXOP in Equations \eqref{eq:e2edlyAMTXOP}. The simulation and analytical results reveal that the model and the simulation agree for both video sequences with acceptable difference up to maximum 10\% which caused by the disregarded overheads. The improvement achieved by ATXOP over HCCA can be straightforwardly implied from Equation \eqref{eq:SIdlyAXTOP} where the unspent transmission time of a station will be available to the next stations in the polling list which in turn schedule the station. It is worth noting that ATXOP and AMTXOP shows superior enhancement with high-quality video, Figure \ref{fig:analyticalhigh}, compared to HCCA as the TXOP duration in the latter is computed for larger MSDU size and subsequently the unspent TXOP duration $T^{i}_{u}$ might be high as well. The effect of multi-polling scheme in enhancing the packet delay in the SI of any $QSTA_i$ is justified by the minimizing the overhead by $(i-1) \times T_{poll}$ in Equation~\eqref{eq:SIdlyAMXTOP}.

\subsection{Performance Analysis over Error-Prone Wireless networks}

Aiming to measure the performance consistency of the proposed mechanism in error-prone wireless networks, we tested the performance with increasing Packet Error Rate (PER) in the system from 1\% to 9\%. In \cite{lasowski2011,einhaus2008} the maximum PER measured is 10\%, we exclude the value of PER that is greater than 10\% as it is unsuitable for most applications \cite{choudhury2008}. We have chosen the case of 12 QSTAs transmitting to a QAP as a worst case scenario in these experiments. The PER was applied to the packets of the uplink traffics. In spite of the fact that the HC assigns the TXOPs based on information piggybacked in each frame, in the case of packet loss the TXOP will be computed as in the HCCA for one or more SI until the reception happen where the HC resume the Dynamic mechanism.  Mean end-to-end delay and aggregate throughput metrics have been measured and the results are discussed as follows.

\begin{enumerate}
\item \textit{End-to-End delay analysis}

The results reveal that the proposed mechanism remains stable minimizing the delay even with increasing amount of packet loss. One can notice that, the increase of End-to-End delay of the packets in High-quality videos, \ref{fig:BERe2eDlyHigh1} and \ref{fig:BERe2eDlyHigh2}, compared to that of the Low-quality ones, \ref{fig:BERe2eDlyLow1} and \ref{fig:BERe2eDlyLow2}, for the same reasons discussed in section \ref{sec:e2edly}.

\begin{figure}
\centering
\includegraphics[scale=0.31,angle=-90]{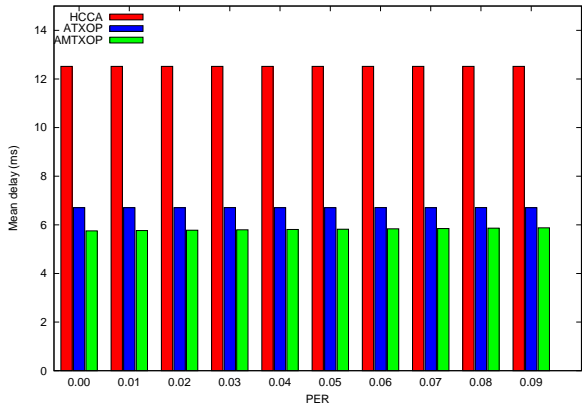}
\caption{Mean Delay with Different PERs for Low-quality Jurassic Park 1}
\label{fig:BERe2eDlyLow1}
\end{figure}

\begin{figure}
\centering
\includegraphics[scale=0.31,angle=-90]{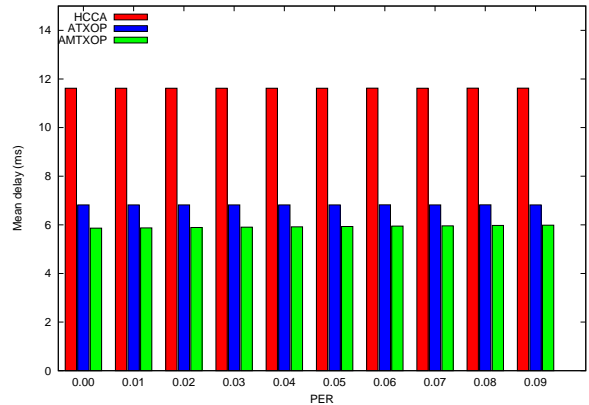}
\caption{Mean Delay with Different PERs for Low-quality Formula 1}
\label{fig:BERe2eDlyLow2}
\end{figure}

\begin{figure}
\centering
\includegraphics[scale=0.31,angle=-90]{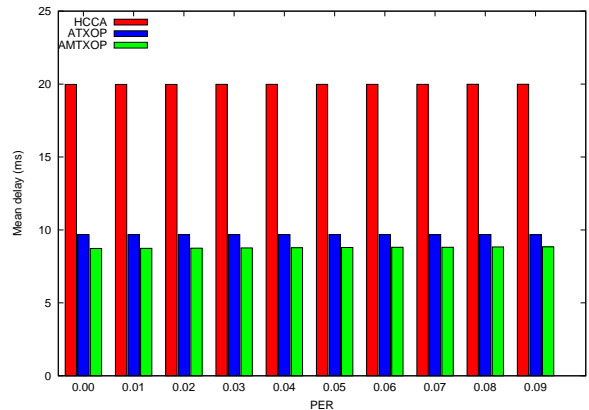}
\caption{Mean Delay with Different PERs for High-quality Jurassic Park 1}
\label{fig:BERe2eDlyHigh1}
\end{figure}

\begin{figure}
\centering
\includegraphics[scale=0.31,angle=-90]{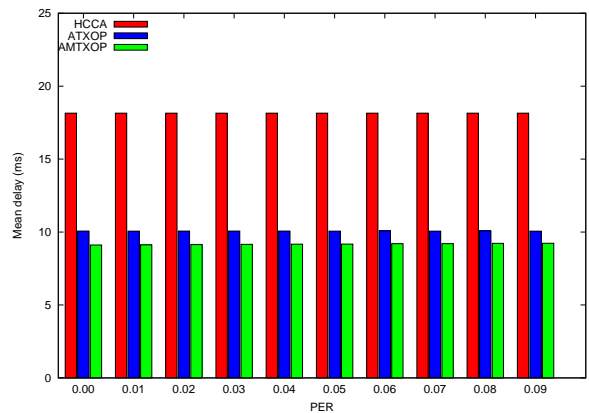}
\caption{Mean Delay with Different PERs for High-quality Formula 1}
\label{fig:BERe2eDlyHigh2}
\end{figure}

\item \textit{Throughput analysis}

The throughput achieved by the proposed mechanism by the presence of packet loss is depicted in Figure \ref{fig:BERe2eThrpHigh3}. The proposed mechanism and HCCA perform similarly attaining the same throughput. Indeed, the achievable throughput is degrading as the PER increases due to the increase of packet loss on the system. 

\begin{figure}
\centering
\includegraphics[scale=0.31,angle=-90]{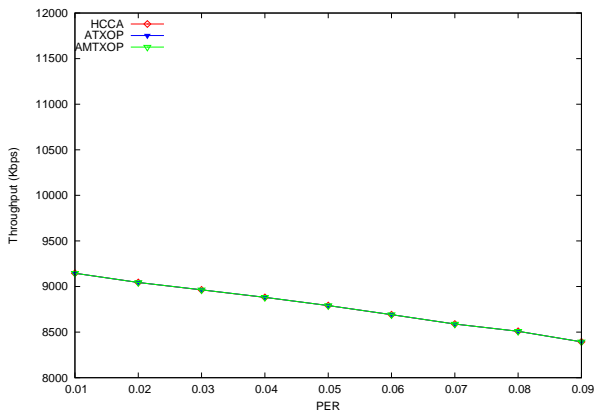}
\caption{Throughput with Different PERs for High-quality Jurassic Park 1}
\label{fig:BERe2eThrpHigh3}
\end{figure}

\end{enumerate}
\subsection{Performance Analysis with Mobility}
In order to investigate the effect of the mobility on the performance of the examined schemes, we used similar topology to that discussed in Section \ref{sec:simSet} except that the stations have been placed to be moved within four diameters each with corresponding PHY rate setting as in Table \ref{tab:mobScen}. It is worth noting that in scenarios a, b and c, the stations experience higher data rate as they move towards the AP and vice versa. These scenarios have been investigated with different movement speeds, namely 5 and 30 meter per second.

\begin{table}[!h]
\centering
\caption {Mobility Scenarios with different PHY rate ranges}
\begin{tabular}{cccc}
\hline
Scenario 	& PHY rate 	& Distance from AP	\\
			& (Mbps)	& (feet)			\\ \hline
a			& 6			& 325 				\\
b			& 18		& 250 				\\
c			& 36		& 200 				\\
d			& 54		& 80 				\\ \hline
\end{tabular}
\label{tab:mobScen}
\end{table}

\begin{figure}
\centering
\subfigure[HCCA]{
\label{fig:Mobility1HCCA} 
\includegraphics[width=.7\linewidth,angle=-90]{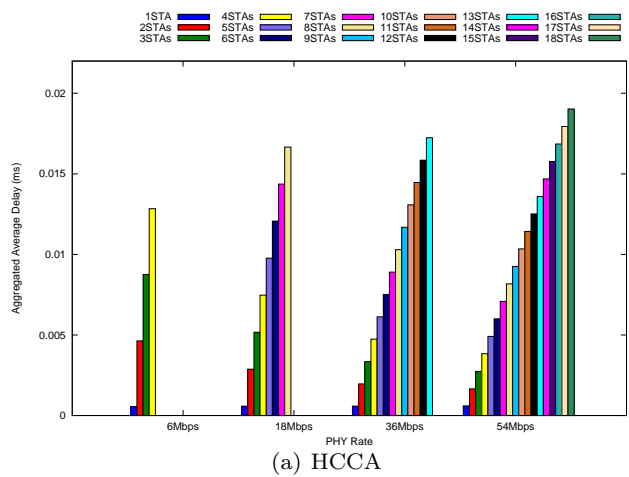}
}
\subfigure[AXTOP]{
\label{fig:Mobility1ATXOP}
\includegraphics[width=.7\linewidth,angle=-90]{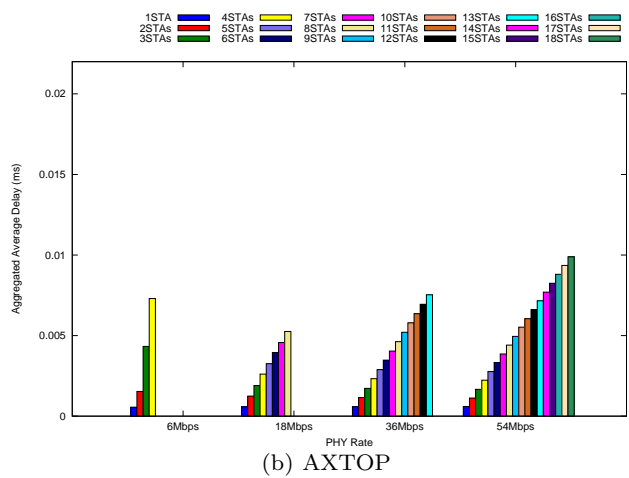}
}
\subfigure[AMXTOP]{
\label{fig:Mobility1AMTXOP}
\includegraphics[width=.7\linewidth,angle=-90]{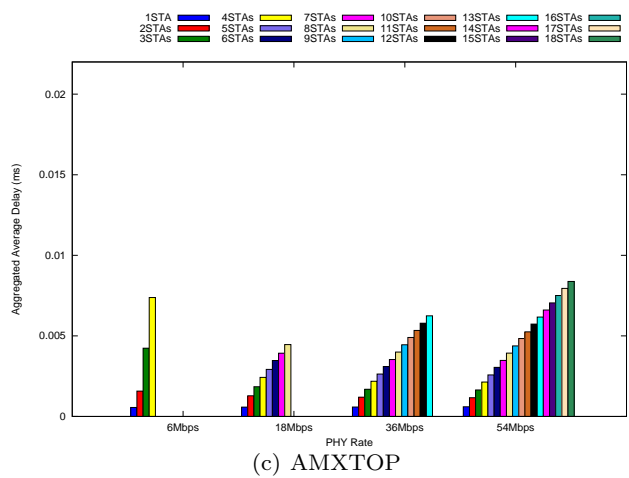}
}
\caption{Delay with Mobility of speed 5m/sec}
\label{fig:Mobility5msec}
\end{figure}

\begin{figure}
\centering
\subfigure[HCCA]{
\label{fig:Mobility2HCCA}
\includegraphics[width=.7\linewidth,angle=-90]{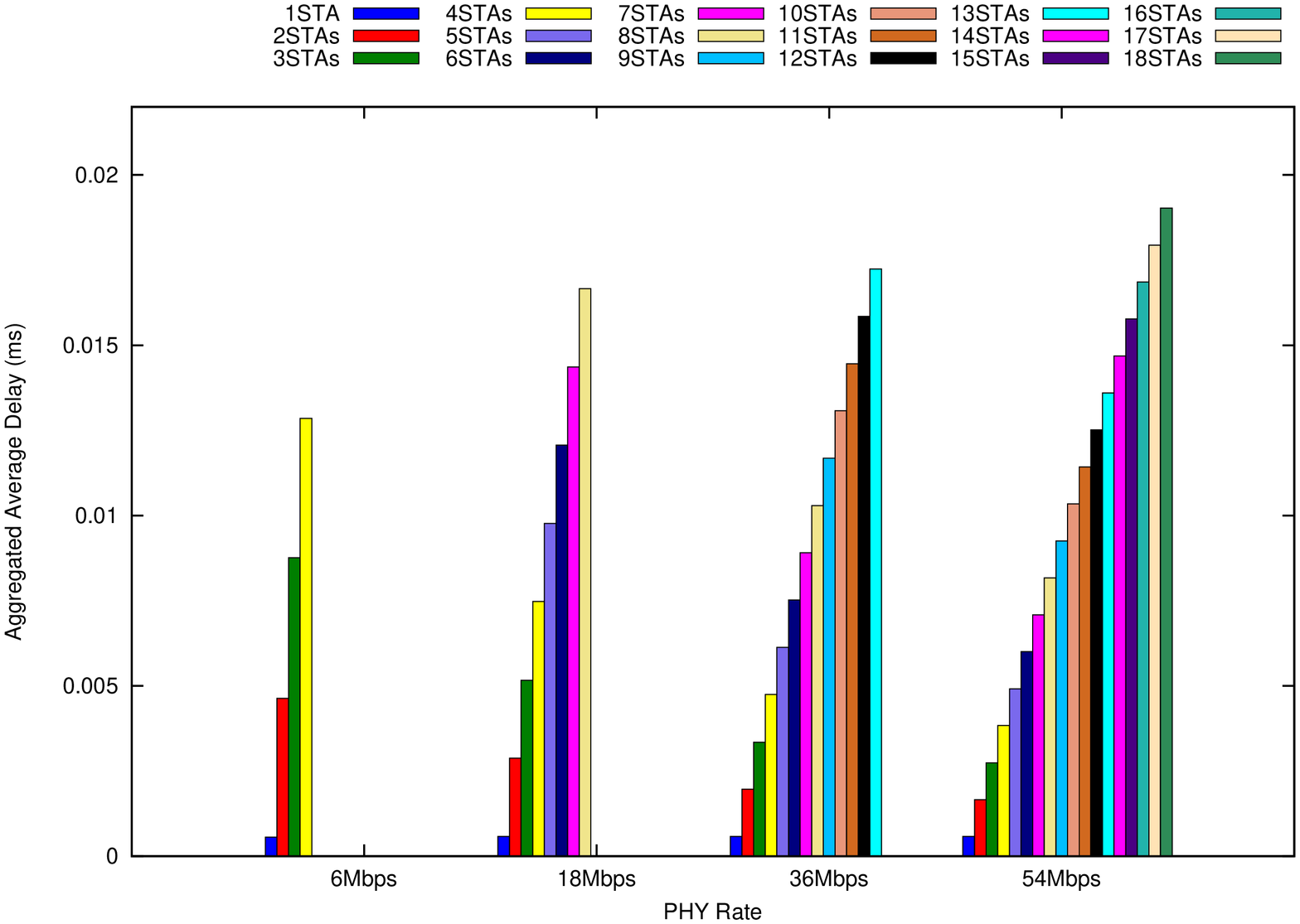}
}
\subfigure[AXTOP]{
\label{fig:Mobility2ATXOP}
\includegraphics[width=.7\linewidth,angle=-90]{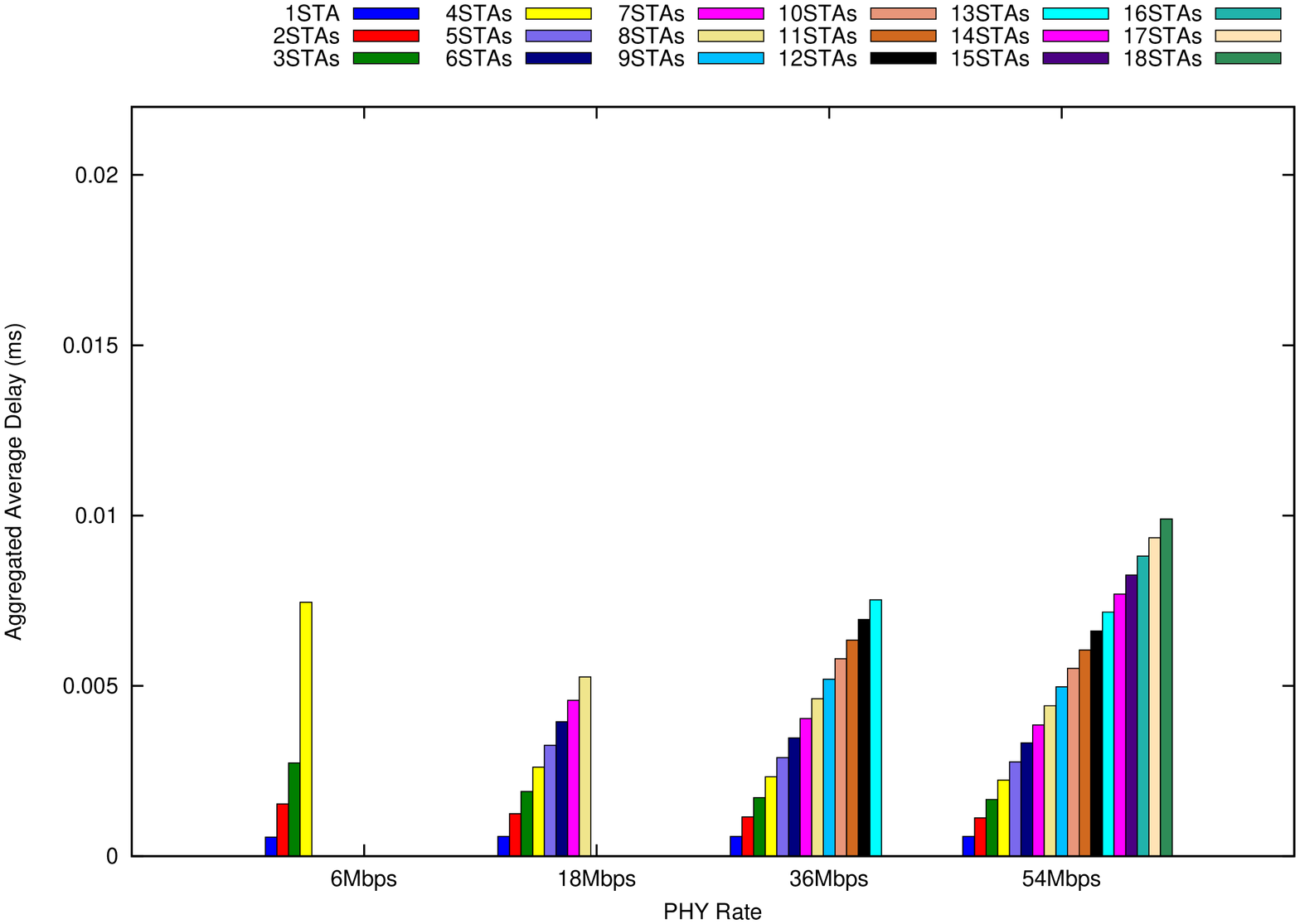}
}
\subfigure[AMXTOP]{
\label{fig:Mobility2AMTXOP}
\includegraphics[width=.7\linewidth,angle=-90]{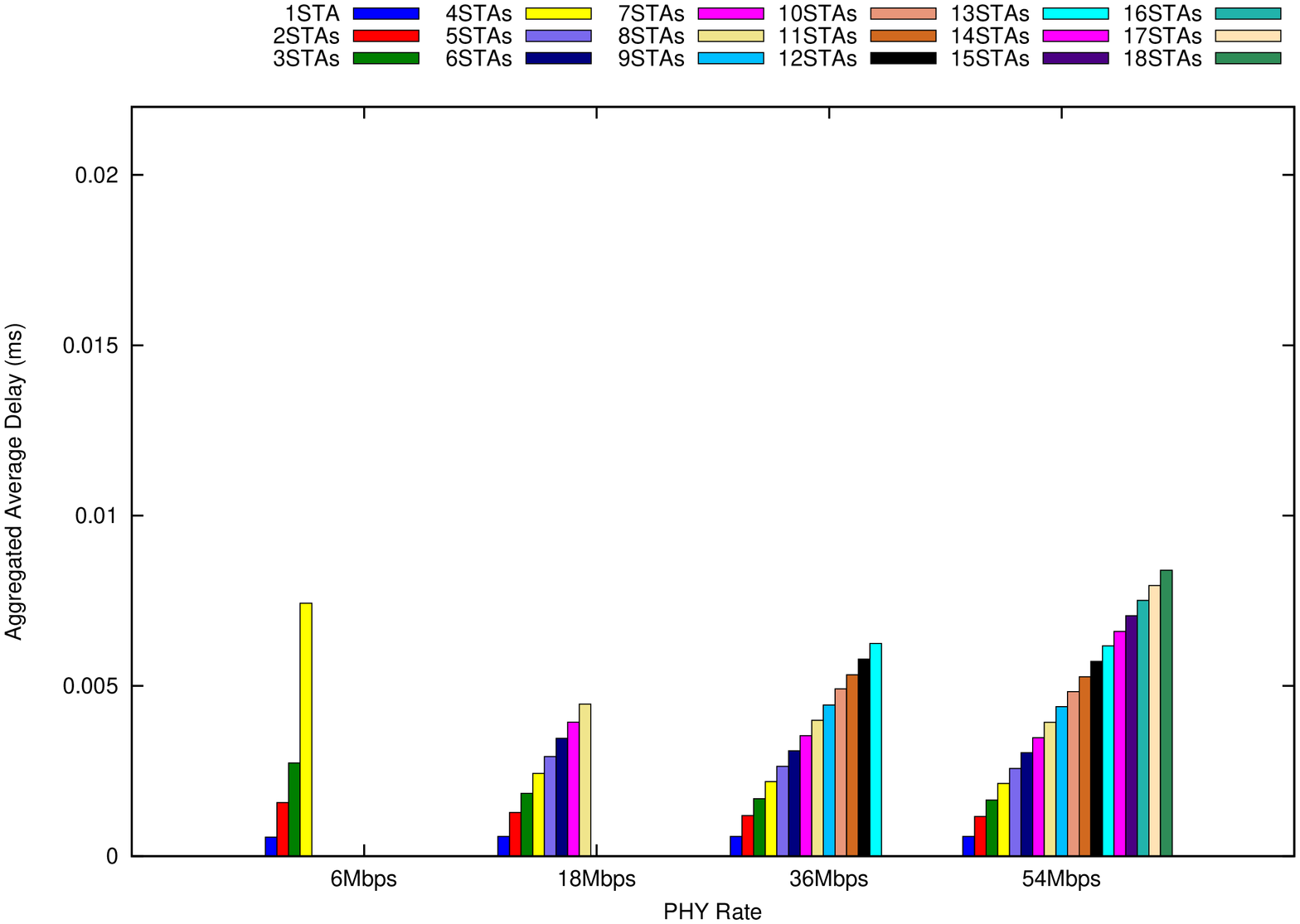}
}
\caption{Delay with Mobility of speed 30m/sec}
\label{fig:Mobility30msec}
\end{figure}

The results in Figure \ref{fig:Mobility1AMTXOP} reveals the robustness of the AMTXOP compared to ATXOP and HCCA (Figures \ref{fig:Mobility1HCCA} and \ref{fig:Mobility1ATXOP} to minimizing the access of each TS using multi-poling scheme. The HCCA shows high delay due to its deficiency to adapt to the real time changes in which the TXOP is computed as a fixed duration at the traffic setup time and used during the traffic life regardless to the actual need of the traffic. Increasing the mobility speed to 30m/sec as illustrated in Figure \ref{fig:Mobility30msec} shows marginally difference compared  to that in Figure \ref{fig:Mobility5msec} as the destination (AP) is only one hop far from the source. One can notice the effect of increasing the PHY on the number of admitted flows which were 4, 8, 13 and 18 TSs in 6Mbps, 18Mbps, 36Mbps and 54Mbps scenarios, respectively.

\subsection{Channel Utilization Analysis}
\begin{figure}
\centering
\includegraphics[scale=0.7]{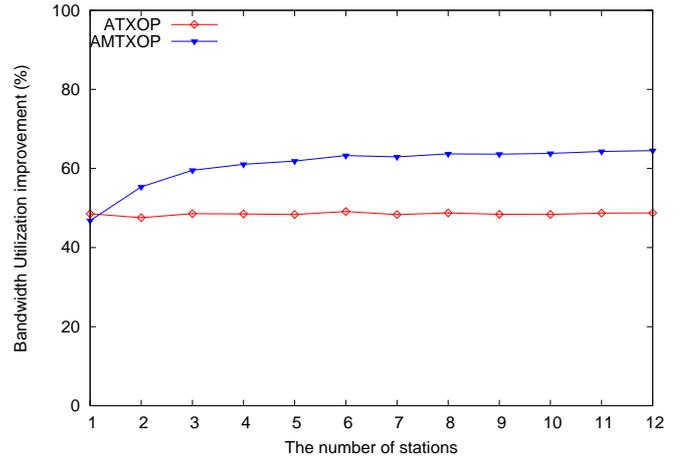}
\caption{Channel Utilization improvement for Low-quality Jurassic Park 1}
\label{fig:ChnlUtillow}
\end{figure}

\begin{figure}
\centering
\includegraphics[scale=0.7]{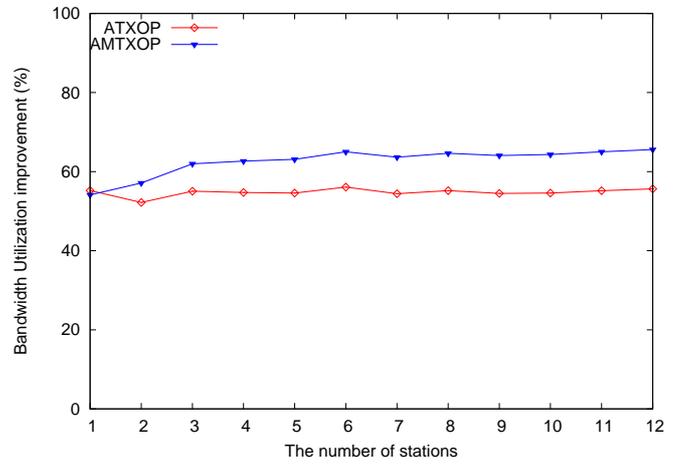}
\caption{Channel Utilization improvement for High-quality Jurassic Park 1}
\label{fig:ChnlUtilhigh}
\end{figure}

In this section, we study the improvement of the bandwidth utilization of the ATXOP and the AMTXOP compared to the HCCA protocol. The Bandwidth utilization improvement is calculated for each $STA_{i}$ in an SI as in Equation \eqref{eq:chnlimpr}

\begin{equation}
\label{eq:chnlimpr}
B_{im}^{i} = \frac{B_{HCCA}^{i}-B_{proposed}^{i}}{B_{HCCA}^{i}}
\end{equation}
where, $B_{HCCA}$ is the channel utilization of HCCA which can be calculated as the proportion of fixed TXOP to the SI, whereas $B_{proposed}$ is for ATXOP and AMTXOP. Figures \ref{fig:ChnlUtillow} and \ref{fig:ChnlUtilhigh} demonstrate the channel utilization improvement as the average of all admitted flows. The ATXOP achieved  utilization improvement over HCCA of about 49\% and 56\% for the low and high quality videos, respectively. Whereas, thanks to multi-polling scheme, the AMTXOP achieves up to 56\% and 66\% over the HCCA.

\section{Conclusion}
\label{sec:conclusion}

In this paper, a novel scheduler called AMTXOP has been  proposed to leverage the performance of the controlled access mode of IEEE 802.11e. The enhancements of AMTXOP is two fold. First, the HC polls QSTAs with regard to fast changing in the traffic profile so as to prevent QSTAs from receiving unnecessary large TXOP. Second, the surplus time of the wireless channel is conserved by reducing the number of poll frames using the multi-polling approach. 

This work shows the significant impact of integrating the multi-polling scheme to the existing piggybacking mechanism in the VBR traffic scheduling. This integrated scheme reduces the packet delay and enhances the channel utilization as well. 

Performance evaluation throughout simulation experiments and analytical study reveals the efficiency of the proposed scheduler over HCCA and ATXOP. Moreover, the schemes were evaluated with mobility scenarios to check the robustness over harsher environments. Thanks to multi-polling concept, the total TXOP duration assigned in AMTXOP was 32\% and 25\% less than that in ATXOP in low and high videos, respectively. This remarkable improvement in the channel utilization has led to significant enhancements in the packet delay. The AMTXOP achieved delay enhancement up to 12\% and 59\% over the ATXOP and HCCA, respectively. Moreover, the gain in the channel utilization is also considered as bonus merit.

\section{Future Directions}
\label{sec:FutureDirection}
In the future, we intend to design an admission control unit (HCCA) to utilize the surplus bandwidth gain using AMTXOP mechanism and manage these bandwidth resources among the HCF functions, EDCA and HCCA for maximizing the number of admitted flows in the system and study the affect of increasing the admitted flows on the TCP flows of the EDCA. Moreover, an enhancement will be introduced to support live video transmission by exploiting the perfecting concept and govern the resulted additional delay budget. The other possible future direction is to enhance the proposed scheduler to cope with more sophisticated wireless conditions and with the MAC level fragmentation and multirate adaptation support.

\section{Acknowledgment} This work was supported by the Malaysian Ministry of High Education under the Fundamental Research Grant Scheme, FRGS/02/01/12/1143/FR.

\begin{thebibliography}{10}
\providecommand{\url}[1]{{#1}}
\providecommand{\urlprefix}{URL }
\expandafter\ifx\csname urlstyle\endcsname\relax
  \providecommand{\doi}[1]{DOI~\discretionary{}{}{}#1}\else
  \providecommand{\doi}{DOI~\discretionary{}{}{}\begingroup
  \urlstyle{rm}\Url}\fi

\bibitem{Abouaissa2013}
{A}bouaissa, A., {B}rahmia, M.e.A., {L}orenz, P.: {I}ncreasing end-to-end
  fairness over {IEEE} 802.11e-based wireless mesh networks.
\newblock {I}nternational {J}ournal of {C}ommunication {S}ystems
  \textbf{26}(1), 1--12 (2013).
\newblock \doi{10.1002/dac.1319}.
\newblock \urlprefix\url{http://dx.doi.org/10.1002/dac.1319}

\bibitem{Ali2013}
{A}li, N.A., {B}ourawy, A., {H}assanein, H.: {S}electivity function scheduler
  for {IEEE} 802.11e {HCCA} access mode.
\newblock {W}ireless {C}ommunications and {M}obile {C}omputing \textbf{13}(2),
  95--110 (2013)

\bibitem{birkos2013}
{B}irkos, K., {T}selios, C., {D}agiuklas, T., {K}otsopoulos, S.: {P}eer
  selection and scheduling of {H}. 264 {SVC} video over wireless networks.
\newblock In: Wireless Communications and Networking Conference (WCNC), 2013
  IEEE, pp. 1633--1638. IEEE (2013)

\bibitem{Cecchetti2012}
{C}ecchetti, G., {R}uscelli, A., {M}astropaolo, A., {L}ipari, G.: {P}roviding
  {V}ariable {TXOP} for {IEEE} 802.11e {HCCA} {R}eal-{T}ime {N}etworks.
\newblock In: Wireless Communications and Networking Conference (WCNC), pp.
  1508--1513 (2012)

\bibitem{cecchettielAL2012}
{C}ecchetti, G., {R}uscelli, A.L., {M}astropaolo, A., {L}ipari, G.: {D}ynamic
  {TXOP} {HCCA} {R}eclaiming {S}cheduler with {T}ransmission {T}ime
  {E}stimation for {IEEE} 802.11e {R}eal-{T}ime {N}etworks.
\newblock In: Proceedings of the 15th ACM international conference on Modeling,
  analysis and simulation of wireless and mobile systems, pp. 239--246. ACM
  (2012)

\bibitem{chaudhuri2013}
{C}haudhuri, S., {R}am, A.: {S}ystem for creating a capsule representation of
  an instructional video (2013).
\newblock US Patent 8,345,990

\bibitem{Chou2011}
{C}hin-{W}en {C}hou, {L}in, K.J., {T}sern-{H}uei {L}ee: {O}n efficient
  multipolling with various service intervals for {IEEE} 802.11e {WLAN}s.
\newblock In: Wireless Communications and Mobile Computing Conference (IWCMC),
  2011 7th International, pp. 1906--1911 (2011)

\bibitem{choudhury2008}
{C}houdhury, S., {G}ibson, J.D.: {T}hroughput optimization for wireless {LAN}s
  in the presence of packet error rate constraints.
\newblock {C}ommunications {L}etters, {IEEE} \textbf{12}(1), 11--13 (2008)

\bibitem{cicconetti2005}
{C}icconetti, C., {L}enzini, L., {M}ingozzi, E., {S}tea, G.: {A} software
  architecture for simulating {IEEE} 802.11e {HCCA}.
\newblock In: {IPS}-{M}o{M}e05: {P}roceeding from the 3rd {W}orkshop on
  {I}nternet {P}erformance, {S}imulation, {M}onitoring and {M}easurement, pp.
  97--104 (2005)

\bibitem{lo2007}
{L}o {C}igno, R., {P}alopoli, L., {C}olombo, A.: {A}nalysis of different
  scheduling strategies in 802.11 e networks with multi-class traffic.
\newblock In: Local Computer Networks, 2007. LCN 2007. 32nd IEEE Conference on,
  pp. 455--462. IEEE (2007)

\bibitem{einhaus2008}
{E}inhaus, M., {K}lein, O., {W}alke, B.: {C}omparison of {OFDMA} resource
  scheduling strategies with fair allocation of capacity.
\newblock In: Consumer Communications and Networking Conference, 2008. CCNC
  2008. 5th IEEE, pp. 407--411. IEEE (2008)

\bibitem{Fang2006raey}
{F}ang, Z., {X}u, S., {W}an, C., {W}ang, Z., {W}u, S., {Z}eng, W.: {M}odeling
  {MPEG}-4 {VBR} {V}ideo {T}raffic by {U}sing {ANFIS}.
\newblock In: D.S. {H}uang, K.~{L}i, G.~{I}rwin (eds.) Intelligent Computing in
  Signal Processing and Pattern Recognition, \emph{Lecture Notes in Control and
  Information Sciences}, vol. 345, pp. 958--963. {S}pringer {B}erlin
  {H}eidelberg (2006)

\bibitem{Fitzek2000}
{F}itzek., F., {R}eisslein, M.: {MPEG-4 and H.263 Video Traces for Network
  Performance Evaluation}.
\newblock Tech. Rep. TKN-00-006, Telecommunication Networks Group, Technische
  Universit{\"a}t Berlin (2000)

\bibitem{Fitzek2001}
{F}itzek, F., {R}eisslein, M.: {MPEG}--4 and {H}.263 {V}ideo {T}races for
  {N}etwork {P}erformance {E}valuation.
\newblock {IEEE} {N}etwork \textbf{15}(6), 40--54 (2001)

\bibitem{Fu2003}
{F}u, H., {Z}hang, L.: {V}ariable segmentation based on intrinsic video rate
  characteristics to transport pre-stored video across networks.
\newblock {I}nternational {J}ournal of {C}ommunication {S}ystems
  \textbf{16}(10), 923--934 (2003).
\newblock \doi{10.1002/dac.629}.
\newblock \urlprefix\url{http://dx.doi.org/10.1002/dac.629}

\bibitem{Gautam2013}
{G}autam, N., {P}etander, H., {N}oel, J.: {A} {C}omparison of the {C}ost and
  {E}nergy {E}fficiency of {P}refetching and {S}treaming of {M}obile {V}ideo.
\newblock In: Proceedings of the 5th Workshop on Mobile Video, MoVid '13, pp.
  7--12. {ACM}, New York, NY, USA (2013)

\bibitem{Grilo2003}
{G}rilo, A., {M}acedo, M., {N}unes, M.: {A} scheduling algorithm for {Q}o{S}
  support in {IEEE}802.11 networks.
\newblock {W}ireless {C}ommunications, {IEEE} \textbf{10}(3), 36 -- 43 (2003)

\bibitem{Haddad2012}
{R}ami {H}addad, {M}ichael {P}.~{M}c{G}arry: {F}eed {F}orward {B}andwidth
  {I}ndication ({FFBI}): {C}ooperation for an accurate bandwidth forecast.
\newblock {C}omputer {C}ommunications \textbf{35}(6), 748 -- 758 (2012)

\bibitem{He2011}
{Y}ong {H}e, {X}iaojun {M}a: {D}eterministic {B}ackoff: {T}oward {E}fficient
  {P}olling for {IEEE} 802.11e {HCCA} in {W}ireless {H}ome {N}etworks.
\newblock {M}obile {C}omputing, {IEEE} {T}ransactions on \textbf{10}(12), 1726
  --1740 (2011)

\bibitem{Huang2010}
{J}eng-{J}i {H}uang, {Y}eh-{H}orng {C}hen, {S}hiung, D.: {A} four-way-polling
  {Q}o{S} scheduler for {IEEE} 802.11e {HCCA}.
\newblock In: {TENCON} 2010 - 2010 {IEEE} {R}egion 10 {C}onference, pp. 1986
  --1991 (2010)

\bibitem{IEEEStand1999}
{IEEE 802.11}: {IEEE} {S}tandard for {I}nformation {T}echnology-
  {T}elecommunications and {I}nformation {E}xchange {B}etween {S}ystems-
  {L}ocal and {M}etropolitan {A}rea {N}etworks- {S}pecific {R}equirements-
  {P}art 11: {W}ireless {LAN} {M}edium {A}ccess {C}ontrol ({MAC}) and
  {P}hysical {L}ayer ({PHY}) {S}pecifications.
\newblock {ANSI}/{IEEE} {S}td 802.11, 1999 {E}dition ({R}2003) pp. i--513
  (1999)

\bibitem{IEEEStandard2007}
{IEEE 802.11e}: {IEEE} {S}tandard for {I}nformation {T}echnology -
  {T}elecommunications and {I}nformation {E}xchange {B}etween {S}ystems -
  {L}ocal and {M}etropolitan {A}rea {N}etworks-{S}pecific {R}equirements-{P}art
  11: {W}ireless {LAN} {M}edium {A}ccess {C}ontrol ({MAC}) and {P}hysical
  {L}ayer ({PHY}) {S}pecifications.
\newblock {IEEE} {S}td 802.11-2007 ({R}evision of {IEEE} {S}td 802.11-1999) pp.
  1 --1076 (2007)

\bibitem{IEEEStandard2012}
{IEEE 802.11e}: {IEEE} {S}tandard for {I}nformation
  technology--{T}elecommunications and information exchange between systems
  {L}ocal and metropolitan area networks--{S}pecific requirements {P}art 11:
  {W}ireless {LAN} {M}edium {A}ccess {C}ontrol {(MAC)} and {P}hysical {L}ayer
  {(PHY)} {S}pecifications.
\newblock {IEEE} {S}td 802.11-2012 ({R}evision of {IEEE} {S}td 802.11-2007) pp.
  1--2793 (2012).
\newblock \doi{10.1109/IEEESTD.2012.6178212}

\bibitem{indexglobal}
{I}ndex, C.V.N.: {G}lobal {M}obile {D}ata {T}raffic {F}orecast {U}pdate,
  2012--2017, {C}isco {W}hite {P}aper, {F}eb. 6, 2013

\bibitem{NSBook2012}
{I}ssariyakul, T., {H}ossain, E.: {I}ntroduction to {N}etwork {S}imulator
  {NS}2.
\newblock {S}pringer (2012)

\bibitem{G7111988}
{ITU}-{T}: {ITU-T} {R}ecommendation {G}.711. {P}ulse {C}ode {M}odulation
  ({PCM}) of {V}oice {F}requencies (1988)

\bibitem{Jansang2011}
{J}ansang, A., {P}honphoem, A.: {A}djustable {TXOP} mechanism for supporting
  video transmission in {IEEE} 802.11e {HCCA}.
\newblock {EURASIP} {J}ournal on {W}ireless {C}ommunications and {N}etworking
  \textbf{2011}(1), 1--16 (2011)

\bibitem{Ammar2011}
{A}l {J}ubari, A.M., {O}thman, M., {A}li, B.M., {H}amid, N.A.W.A.: {TCP}
  performance in multi-hop wireless ad hoc networks: challenges and solution.
\newblock {EURASIP} {J}ournal on {W}ireless {C}ommunications and {N}etworking
  \textbf{2011}(1), 1--25 (2011)

\bibitem{ByungSeoKim2005}
{B}yung-{S}eo {K}im, {S}ung~{W}on {K}im, {Y}uguang {F}ang, {W}ong, T.:
  {T}wo-step multipolling {MAC} protocol for wireless {LAN}s.
\newblock {S}elected {A}reas in {C}ommunications, {IEEE} {J}ournal on
  \textbf{23}(6), 1276--1286 (2005)

\bibitem{koenen1999}
{K}oenen, R.: {MPEG-4} {M}ultimedia for our {T}ime.
\newblock {S}pectrum, {IEEE} \textbf{36}(2), 26--33 (1999)

\bibitem{Koenen2002}
{K}oenen, R.: {O}verview of the {MPEG-4} {S}tandard.
\newblock {ISO}/{IEC} {JTC}1/{SC}29/{WG}11 {N} \textbf{1730}, 11--13 (2002)

\bibitem{lasowski2011}
{L}asowski, R., {G}schwandtner, F., {S}cheuermann, C., {D}uchon, M.: {A} multi
  channel synchronization approach in dual radio vehicular ad-hoc networks.
\newblock In: Vehicular Technology Conference (VTC Spring), 2011 IEEE 73rd, pp.
  1--5. IEEE (2011)

\bibitem{Lee2009}
{L}ee, D.Y., {K}im, S.R., {L}ee, C.W.: {A}n {E}nhanced {EDD} {Q}o{S}
  {S}cheduler for {IEEE} 802.11e {WLAN}.
\newblock In: Advances in Computational Science and Engineering,
  \emph{Communications in Computer and Information Science}, vol.~28, pp.
  45--59. {S}pringer {B}erlin {H}eidelberg (2009)

\bibitem{Lee2013}
{L}ee, T.H., {H}uang, Y.W.: {Q}uality of service guarantee for real-time {VBR}
  traffic flows with different delay bound and loss probability requirements in
  {WLAN}s.
\newblock {J}ournal of the {C}hinese {I}nstitute of {E}ngineers \textbf{36}(4),
  471--487 (2013)

\bibitem{Mao2007}
{M}ao, G., {L}iu, H.: {R}eal {T}ime {V}ariable {B}it {R}ate {V}ideo {T}raffic
  {P}rediction: {R}esearch {A}rticles.
\newblock {I}nt. {J}. {C}ommun. {S}yst. \textbf{20}(4), 491--505 (2007)

\bibitem{Almaqri2013}
{A}l {M}aqri, M., {O}thman, M., {A}li, B., {H}anapi, Z.: {A}daptive {TXOP}
  assignment for {Q}o{S} support of video traffic in {IEEE} 802.11e networks.
\newblock In: RF and Microwave Conference (RFM), 2013 IEEE International, pp.
  144--149 (2013)

\bibitem{NS2}
{M}c{C}anne, S., {F}loyd, S.: {NS} network simulator (1995)

\bibitem{Panos2002}
{N}asiopoulos, P., {W}ard, R.K.: {E}ffective multi-program broadcasting of
  prerecorded video using {VBR} {MPEG}-2 coding.
\newblock {B}roadcasting, {IEEE} {T}ransactions on \textbf{48}(3), 207--214
  (2002)

\bibitem{navarro2013}
{N}avarro {O}rtiz, J., {A}meigeiras, P., {R}amos {M}unoz, J.J., {L}opez
  {S}oler, J.: {R}emoving redundant {TCP} functionalities in wired-cum-wireless
  networks with {IEEE} 802.11e {HCCA} support.
\newblock {I}nternational {J}ournal of {C}ommunication {S}ystems  (2013).
\newblock \doi{10.1002/dac.2546}

\bibitem{rashid2008}
{R}ashid, M., {H}ossain, E., {B}hargava, V.: {C}ontrolled {C}hannel {A}ccess
  {S}cheduling for {G}uaranteed {Q}o{S} in 802.11e-{B}ased {WLAN}s.
\newblock {W}ireless {C}ommunications, {IEEE} {T}ransactions on \textbf{7}(4),
  1287 --1297 (2008)

\bibitem{ruscelli2013}
{A}nna~{L}ina {R}uscelli, {G}abriele {C}ecchetti, {A}ngelo {A}lifano,
  {G}iuseppe {L}ipari: {E}nhancement of {Q}o{S} support of {HCCA} schedulers
  using {EDCA} function in {IEEE} 802.11e networks.
\newblock {A}d {H}oc {N}etworks \textbf{10}(2), 147 -- 161 (2012)

\bibitem{ruscelli2014}
{R}uscelli, A.L., {C}ecchetti, G.: {A} {IEEE} 802.11 e {HCCA} {S}cheduler with
  a {R}eclaiming {M}echanism for {M}ultimedia {A}pplications.
\newblock {A}dvances in {M}ultimedia \textbf{2014} (2014)

\bibitem{ruscelli2011}
{R}uscelli, A.L., {C}ecchetti, G., {M}astropaolo, A., {L}ipari, G.: {A} greedy
  reclaiming scheduler for {IEEE} 802.11 e {HCCA} real-time networks.
\newblock In: Proceedings of the 14th ACM international conference on Modeling,
  analysis and simulation of wireless and mobile systems, pp. 223--230. ACM
  (2011)

\bibitem{MPEG11997}
{S}ikora, T.: {MPEG} {D}igital {V}ideo-{C}oding {S}tandards.
\newblock {IEEE} {S}ignal {P}rocessing {M}agazine \textbf{14}(5), 82--100
  (1997)

\bibitem{soares1998}
{S}oares, L.D., {P}ereira, F.: {MPEG-4}: {A} {F}lexible {C}oding {S}tandard for
  the {E}merging {M}obile {M}ultimedia applications.
\newblock In: Personal, Indoor and Mobile Radio Communications, 1998. The Ninth
  IEEE International Symposium on, vol.~3, pp. 1335--1339 (1998)

\bibitem{Teixeira2013}
{T}eixeira, M.A., {G}uardieiro, P.R.: {A}daptive packet scheduling for the
  uplink traffic in {IEEE} 802.16e networks.
\newblock {I}nternational {J}ournal of {C}ommunication {S}ystems
  \textbf{26}(8), 1038--1053 (2013).
\newblock \doi{10.1002/dac.1390}.
\newblock \urlprefix\url{http://dx.doi.org/10.1002/dac.1390}

\bibitem{Zhang2013}
{B}ing {ZHANG}, {M}ao-de {MA}, {C}hun-feng {LIU}, {Y}an-tai {SHU}:
  {I}mprovement of polling and scheduling scheme for real-time transmission
  with \{HCCA\} of \{IEEE\} 802.11p protocol.
\newblock {T}he {J}ournal of {C}hina {U}niversities of {P}osts and
  {T}elecommunications \textbf{20}(3), 60 -- 66 (2013)

\end{thebibliography}
\end{document}